\documentclass[12pt]{JHEP3}
\usepackage{mathrsfs}
\usepackage{amsmath,amssymb}
\usepackage{epsfig}
\usepackage{relsize}

\input epsf

\def\x'{\mathaccent 19 x}
\def\y'{\mathaccent 19 y}
\def\n'{\mathaccent 19 n}
\def\u'{\mathaccent 19 u}

\def\et'{\mathaccent 19 \eta}
\def\th'{\mathaccent 19 \theta}
\def\lam'{\mathaccent 19 \lambda}
\def\varet'{\mathaccent 19 \vartheta}
\def\rh'{\mathaccent 19 \rho}
\def\ph'{\mathaccent 19 \phi}
\def\xb'{\mathaccent 19 {\bar{x}}}

\newcommand{\PP}{\mathcal P}

\def\tl{\widetilde{\lambda}}

\def\l{{\lambda}}

\def\N{${\cal N}=4$ }

\def\sl(2){\alg{sl}(2)}
\def \PP {{\cal P}}
\def\det{\hbox{det}}
\def\be{\begin{equation}}
\def\ee{\end{equation}}

\newcommand{\bea}{\begin{eqnarray}}
\newcommand{\eea}{\end{eqnarray}}
\newcommand\bp{\hbox{\larger\larger $\pi$}}

\def\GS{Green-Schwarz }

\def\tpd{\theta'^\dagger}
\def\etapd{\eta'^\dagger}
\def\etad{\eta^\dagger}
\def\td{\theta^\dagger}
\def\T{\Theta}
\def\Ts{\Theta_*}
\def\Td{\Theta^\dagger}
\def\Tdt{\Theta^{\dagger,t}}
\def\Tpd{\Theta'^\dagger}
\def\Tpdt{\Theta'^{\dagger,t}}

\def\a {\alpha}
\def\b {\beta}
\def\s {\sigma}
\def\pa {\partial}

\def\g {\gamma}
\def\om {\omega}
\def\p{\phi}
\def\la{\label}

\def\ov{\over}
\def\tr{{\rm tr}}

\def\str{\text{Str}}
\def\Tr{\text{Tr}}
\def\G{\Gamma}
\def\S{\Sigma}
\def\H{{\cal H}}

\newcommand{\alg}[1]{\mathfrak{#1}}
\newcommand{\su}{\alg{su}}
\newcommand{\so}{\alg{so}}

\newcommand{\sign}{\alg{s}}
\newcommand{\psu}{\alg{psu}}
\newcommand{\un}{\alg{u}}

\newcommand{\AdS}{{\rm  AdS}_5\times {\rm S}^5}
\newcommand{\ads}{{\rm  AdS}_5\times {\rm S}^5}

\def\L{\mathscr L}

\author{Sergey Frolov$^{a}$\footnote{e-mail: frolovs@aei.mpg.de, plefka@physik.hu-berlin.de, marzam@aei.mpg.de} \footnote{ Also at SUNYIT, Utica, USA and Steklov
Mathematical Institute, Moscow.}, Jan Plefka$^{b}$ and Marija
Zamaklar$^{a}$ \\ $^{a}$ {\it Max-Planck-Institut f\"ur
Gravitationsphysik, Albert-Einstein-Institut}\\ ~~Am M\"uhlenberg 1,
D-14476 Potsdam, Germany\\ $^{b}$ {\it Humboldt-Universit\"at zu
Berlin, Institut f\"ur Physik,\\ ~~Newtonstra{\ss}e 15, D-12489
Berlin, Germany}\\} 
\abstract{ We use the uniform light-cone gauge to derive an
  \emph{exact} gauge-fixed Lagrangian and light-cone Hamiltonian for
  the Green-Schwarz superstring in $\ads$.  We then quantize the
  theory perturbatively in the near plane-wave limit, and compute the
  leading $1/J$ correction to a generic string state from the rank-1 subsectors.  These investigations enable us to propose a new set of
  light-cone Bethe equations for the quantum string.  
  The equations have a simple form and yield the
  correct spinning string and flat space limits.  Finally, we clarify
  the notion of closed sectors in string theory by proving the
  existence of perturbative effective string Hamiltonians which are
  direct analogues of (all loop) dilatation operators in the dual
  ${\cal N}=4$ gauge theory. }

\title{The \mathversion{bold} $\AdS$ \mathversion{normal} Superstring
in Light-Cone Gauge and its Bethe Equations}
\preprint{AEI-2006-011\\HU-EP-06/08}
\begin{document}

\renewcommand{\thefootnote}{\arabic{footnote}}
\setcounter{footnote}{0}

\section{Introduction}

The quantization of the Green-Schwarz superstring
\cite{Metsaev:1998it} in the $\AdS_5$ background and the determination
of its quantum spectrum continues to be one of the great challenges in
string theory. The $\AdS$ space-time background is distinguished by
the fact that it constitutes one of the three maximally supersymmetric
solutions of ten dimensional type IIB supergravity \cite{AdS5}, with the other two
being flat Minkowski space and the IIB plane-wave geometry
\cite{Blau:2001ne}, connected to $\AdS$ through suitable
Penrose limits \cite{Blau:2002dy}.  While the superstring spectrum in
the latter two backgrounds is straightforwardly attainable \cite{Metsaev:2001bj} (at least
in a light-cone gauge) the situation is considerably more involved for
$\AdS$. Next to these structural issues, the greatest interest in the
$\AdS$ quantum string stems from the AdS/CFT duality conjecture
\cite{AdSCFT,Gubser:1998bc,Witten}, asserting the equivalence of the
string spectrum to the spectrum of scaling dimensions of composite,
gauge invariant operators in the dual ${\cal N}=4$ $U(N_c)$ super
Yang-Mills theory in the large $N_c$ limit.

Here very important progress has been made during recent years
building on the key concept of integrability\footnote{For a list of
reviews see \cite{revs}.}. On the gauge theory side it has been
established that in planar perturbation theory the dilatation operator
\cite{Beisert:2003tq}, whose spectrum yields the desired scaling
dimensions of composite, gauge invariant operators, is isomorphic to
the Hamiltonian of an {\it integrable} quantum spin chain
\cite{Minahan:2002ve}.\footnote{While integrability is know to be
broken beyond the planar level, it seem to be preserved if one is
focusing on the specific set of most probable string splitting channels \cite{split}.}
Here integrability secures the existence of a Bethe ansatz which
enables one to reformulate the quantum spectral problem into the
solution of a set of non-linear algebraic equations, the Bethe
equations. These insights, at present firmly established up to the
three loop order in the 't Hooft coupling $\lambda:=g_{\rm YM}\,
N^2_c$ \cite{Beisertsu23, Eden:2005ta} in certain closed subsectors of
the full $PSU(2,2|4)$ symmetry group, led to the formulation of an
exciting conjecture on the {\it all}-loop structure of these Bethe
equations \cite{BDS}.  This conjecture has been by now extended to the
full $PSU(2,2|4)$ case in the gauge theory \cite{BSsu12} and is
believed to hold in an asymptotic sense, where the classical scaling
dimension of the operator in question determines the loop order to
which a prediction is made by the Bethe equations. Moreover, in the
prominent minimal compact, bosonic subsector of $SU(2)$ the
conjectured gauge theory Bethe equations \cite{BDS} were recently
shown to arise microscopically from the well-known Hubbard model at
half filling \cite{Hubbard}. Whether this surprising connection is
indeed fully realized beyond three loops remains to be seen, for this
a four loop computation on the gauge theory side would have to be
performed.
 
In view of these promising developments on the gauge theory side, it
is clear that we need to increase our knowledge of the quantum string
spectrum as well.

While it is unclear at present how to attack the question of the exact
spectrum, important progress has been made by mapping out the quantum
spectrum perturbatively around limiting, solvable islands within the
unknown full $\AdS$ territory.  One very well studied such island is
the IIB plane-wave background, obtainable as a limit of $\AdS$ when a
single angular momentum $J$ on the ${\rm S}^5$ becomes large
\cite{BMN}. Here the first $1/J$ corrections to the spectrum have been
established in a series of papers by Callan et.~al.~in
\cite{Callan}. A wealth of islands is obtained by applying the
semiclassical approach \cite{GKP2}, and looking at classical spinning
string solutions where several spins on $\AdS$ and angular momenta on
the ${\rm S}^5$ become large \cite{FTspin}.  In this limit the
classical string energies can be shown to provide the leading
contribution to the true quantum spectrum as long as at least one
angular momentum on the ${\rm S}^5$ becomes large. Moreover, for a
number of explicit solutions quantum fluctuation expansions about them
have been performed \cite{fluct}.  A further known but not so well
studied island is the large radius limit of the $\AdS$ geometry
limiting to flat Minkowski space. Here one presently only knows the
leading $\lambda^{1/4}\, \sqrt{|n|}$ behaviour of the spectrum
\cite{Gubser:1998bc}, corrections in the $\lambda\to\infty$ expansion
are unknown territory. 

As a matter of fact all these computations of the quantum  spectrum
in the various limits which allow for an extrapolation to weak coupling
$\lambda\ll 1$ display a worrisome \emph{disagreement} with the
gauge theory results at the three loop order, which might be related
to an order of limits problem \cite{BDS}. Resolving this issue is a pressing
problem for the AdS/CFT conjecture.

The sigma-model describing classical $\AdS$ strings
\cite{Metsaev:1998it} is an integrable model \cite{Bena:2003wd} at the
classical level.  One certainly hopes integrability to persist also in
the quantum theory, although it is unclear at present how this could
be precisely implemented.  Inspired by the all-loop Bethe ansatz
conjectures on the gauge theory side and aware of the obtained data in
the plane wave, flat space and spinning strings limits Arutyunov,
Staudacher and one of the present authors were able to write down a
set of quantum string Bethe equations \cite{AFS} which are
structurally very similar to the gauge theory equations of \cite{BDS},
differing by a so-called dressing factor which depends on (an infinite
set of) undetermined functions of $\lambda$ and thus taking into 
account the three loop discrepancies.  These functions should
be determined by comparison with quantum string data. First steps in
this direction have been performed in
\cite{fluct,ZamaSchaefer,BeisertTseytlin}.  The quantum string Bethe
equations of \cite{AFS} have been also generalized to the full
$PSU(2,2|4)$ setting in \cite{BSsu12} \footnote{An alternative
approach towards the quantum string spectrum not rooted in gauge
theory insights has been pursued in \cite{joe}. Here the focus is
on the existence of the S-matrix.}.

So on the string side two central questions arise: Is the quantum
string spectrum indeed described by this (or a similar) set of Bethe
equations? And: How do such Bethe equations arise at all from a
treatment of the {\it quantum} string?

Addressing these questions is a complicated problem, and the present
paper sets the stage for an extension of the above-mentioned
perturbative studies of the quantum string spectrum by providing a
novel, economic gauge fixed description of the full $\AdS$ system.  We
establish the exact form of the action in the novel uniform light-cone
gauge, recently introduced in a truncation to the $SU(1|1)$ subsector
\cite{AFlc}.  This gauge choice generalizes the phase-space light-cone
gauge of Goddard et.~al.~\cite{Goddard:1973qh} to curved backgrounds.
It amounts to rewriting the string action in a first order form and
introducing the light cone coordinates $x_\pm:=(\phi\pm t)/2$ where
$t$ is the global time coordinate of $AdS_5$ conjugate to the energy
$E$ and $\phi$ is an angle of the ${\rm S}^5$ whose conjugate variable
is the angular momentum $J$.\footnote{Let us mention that in contrast
to flat space where all null-geodesics are equivalent, in 
$\AdS$ there are two inequivalent sets of null geodesics: 
one corresponding to a particle orbiting around the big circle on
${\rm S}^5$, while not moving in ${\rm AdS}_5$ and one where
the particle moves only in the radial direction of ${\rm AdS}_5$ in 
global coordinates (i.e. parallel to the boundary, in the Poincare
coordinates). These two geodesics in turn lead to two possible choices
of the light cone coordinates, and hence to two different light cone
gauge fixings.  The latter coordinates were used in
\cite{Metsaev:2000yu}. This choice is natural from the perspective of
reaching the flat space limit as a Penrose limit around this geodesic
yields flat space. The former coordinates were used in
\cite{Metsaev:2001bj} and are natural for the purposes of reaching
the plane-wave limit.  } The gauge consists of fixing $x_+=\tau$ and
$p_+=P_+={\rm const}$ where $p_+$ is conjugate to $x_-$, along with a
convenient fixing of the local fermionic kappa-symmetry to be
discussed.

We establish an {\it exact} first order form of the superstring action
in this gauge, consisting of a kinetic term which determines the
(complicated) Poisson structure of the theory and an exact world-sheet
Hamiltonian $H_{lc}=-P_-$.  In order to perform a perturbative
quantization of the system about suitable limits one needs to expand
the Lagrangian in the number of physical (transverse) fields.  In
addition one needs to perform a field redefinition, in order to secure
the standard Poisson structure and hence the canonical commutation
relations.  We perform this program explicitly for the case of the
near-plane wave limit, which in the uniform light cone gauge amounts
to taking the $P_+\to\infty$ limit with $\lambda/P_+^2$ held fixed,
and expanding the Langrangian up to the quartic order in transverse
fields.

The resulting world-sheet Hamiltonian $H_{lc}$ and the parameter $P_+$
are related to the global energy $E$ and angular momentum $J$ via
\begin{equation}
\label{EJ}
H_{lc}=-P_-=E-J \qquad P_+=E+J \,.
\end{equation}
Since $H_{lc}$ itself is a function of $P_+$ which one determines in
perturbation theory one obtains an equation $E=J+H_{lc}(E+J)$. Solving
this in turn for $E$ yields the energy $E=E(J)$. This repackaging of
the spectral problem for $E$ appears to be very natural from the
string viewpoint. Indeed one easily establishes, that the first
$1/P_+$ correction in the closed rank one subsectors of the world-sheet
energy $H_{lc}$ is determined by a universal expression multiplied by
$\sign$ with $\sign=\{1,0,-1\}$ for $\{\su(2),\su(1|1),\sl(2)\}$
respectively.

Interestingly, the energy shifts for the $H_{lc}$ eigenvalues also
follow from a rather simple set of {\it light-cone Bethe equations} of
the form
\footnote{Note that the wording ``light cone'' used here does not refer to
the gauge used to derive the worldsheet Hamiltonian.  In the {spinning
string}
limit, the equations (\ref{Blc}) reduce to the integral equations of
\cite{Kazakov:2004qf, Kazakov:2004nh} which are derived from
the gauge unfixed sigma model action.
The main reason why our equations have a different form than those of
\cite{AFS} is because they are diagonalizing a different (but
equivalent) infinite set of charges. For example, the role of the
global energy $E$ is replaced by the light cone energy $-P_-$.  
}

\begin{equation}
\label{Blc}
e^{i\, p_k\,{{P_++\sign\, M}\over 2}} = \prod_{j=1,\, j \neq k}^M 
\left({x_k^+ - x_j^-  \over x_k^- - x_j^+}\, 
\right)^\sign\, ,
\end{equation}
where $p_k$ denote the quasi-momenta, $M$ is the number of string oscillator
excitations and the variables $x^\pm_k=x^\pm(p_k)$ 
are the ones introduced in 
\cite{BDS,Beisert:2004jw} related to the quasi-momenta as 
\begin{equation}
x^{\pm}(p) ={1\over 4}\,\Bigl (\,\cot{p\over 2}\pm i\, \Bigr)\, 
\Bigl (1 + H_{lc}(p) \Bigr ) \, .
\end{equation}
Note that this definition depends on the dispersion relation of the
light-cone system $H_{lc}(p)$.  To the order we have computed (${\cal
O}(1/P_+)$) it is given by $H_{lc}(p) = \sqrt{1 + {\lambda \over
4\pi^2}\,p^2}$.  This dispersion relation will be corrected at higher
orders in the $1/P_+$ expansion. However, the simplest guess for the
all orders structure inspired by the gauge theory Bethe ans\"atze is
clearly
\begin{equation}
\label{eins}
H_{lc}(p) = \sqrt{1 + {\lambda \over \pi^2}\sin^2 \left( {p \over 2}\right)} \, .
\end{equation}
A necessary condition on our light-cone Bethe
equations (\ref{Blc}), is the correct behaviour both in the spinning
string and strong coupling limits. This is explicitly demonstrated in
section 8.  The novel feature of this set of Bethe equations is
that the exponent on the left hand side is not an integer any longer,
as $P_+=E+J$ where $E$ is the global energy we wish to determine in
the end.  Moreover, the dressing factor
which was present for the quantum string equations \cite{AFS} is now
absent. It may appear, however, at higher orders in $1/P_+$.

Much of the material presented in this paper is of a rather technical
nature, which we have tried to delegate to the appendices as much as
possible. In the main text we present the logic and the flow in the
construction and uniform light-cone gauge fixing of the superstring on
$\AdS$. The leading corrections in the near-plane wave expansion are
presented in great detail, along with a discussion of the most
prominent closed subsectors and their energy shifts. Here we reproduce
the results of \cite{Callan} in a very economic fashion.  We also
comment on the emergence of effective Hamiltonians for the closed
subsectors, which are the direct analogue of the dilatation operators
in the parallel subsectors in the gauge theory.  In the final chapter
we present the derivation of the light-cone Bethe equations stated
above and show that its thermodynamic and strong coupling limits are
in agreement with previous results.

\section{Review of the Superstring on $\AdS$}
In this section we review the structure of 
the Green-Schwarz superstring action in the $\AdS$ space-time geometry
following closely the discussion in \cite{Alday:2005gi, Alday:2005jm}. The superstring 
is formulated as a two-dimensional non-linear sigma-model 
whose target space is given by the coset manifold
\cite{Metsaev:1998it} 
{\footnotesize \bea \label{ce}
\frac{\rm PSU(2,2|4)}{{\rm SO(4,1)}\times {\rm SO(5)}} \, . \eea }
The full action is given by the sum of the non-linear
sigma-model action and a topological Wess-Zumino term which is fixed uniquely 
by requiring $\rm PSU(2,2|4)$ and $\kappa$-symmetry invariance.

Let us first discuss some basic facts about the supergroup $\rm PSU(2,2|4)$
that is the isometry group of the $\AdS$ superspace 
and the corresponding Lie superalgebra $\psu(2,2|4)$. 

\subsection{Superalgebra $\psu(2,2|4)$}
A convenient description of the superalgebra $\su(2,2|4)$ is 
provided by $8\times 8$ supermatrices
$M$ which can be written in terms of $4\times 4$ blocks as 
\bea
M=\left(
\begin{array}{cc}
  A & X \\
  Y & D
\end{array} \right)\, .
\eea 
Here the matrices $A$ and $D$ are Grassmann even and $X,Y$ are Grassmann odd.
The superalgebra $\su(2,2|4)$ is 
singled out by 
requiring $M$ to have vanishing supertrace
${\rm str}M={\rm tr}A-{\rm tr}D=0$ and to satisfy the following
reality condition 
\bea 
\label{real} HM+M^{\dagger}H=0\, . 
\eea 
The choice of the hermitian
matrix $H$ is not unique and we choose $H$ to be of the 
diagonal form 
\bea  H=\left(
\begin{array}{rr}
  \Sigma & 0 \\
  0 & \mathbb{I}
\end{array} \right)\, ,
\eea where $\Sigma$ is the following matrix \bea \Sigma=\,
{\footnotesize\left(
\begin{array}{cccc}
  1 & 0 & 0 & 0 \\
  0 & 1 & 0 & 0 \\
   0 & 0 & -1 & 0 \\
   0 & 0 & 0 & -1
\end{array} \right)} \eea
and $\mathbb{I}$ denotes the identity matrix of the corresponding
dimension. This choice of $H$ makes it obvious that the bosonic matrices 
$A$ and $D$ belong to the algebras
$\un(2,2)$ and $\un(4)$ respectively. The condition (\ref{real}) also implies 
that the fermionic matrices $X$ and $Y$ 
are conjugated to each other via the relation
\bea\la{yxd}
Y=-X^{\dagger}\Sigma \,.
\eea
Only the supertraceless combination of the two $\un(1)$ generators of 
$\un(2,2)$ and $\un(4)$ belongs to $\su(2,2|4)$.
It is represented by the anti-hermitian matrix $i\mathbb{I}$.
Thus, the bosonic subalgebra of $\su(2,2|4)$ is 
\bea 
\su(2,2)\oplus \su(4)\oplus \un(1)\, . 
\eea 
The superalgebra $\psu(2,2|4)$ is defined as the
{\it quotient algebra} of $\su(2,2|4)$ over this $\un(1)$ factor. 
It has no realization in terms of $8\times 8$ supermatrices.

\medskip

The construction of the superstring action uses the $\mathbb{Z}_4$ grading
of the superalgebra $\su(2,2|4)$
defined by the automorphism $M\to \Omega(M)$ with 
\bea
\label{Omega} \Omega(M)= \left(
\begin{array}{rr}
  K A^t K ~&~ -K Y^tK \\
   K X^t K ~&~ K D^t K
\end{array} \right)\, ,
 \eea
where $A^t$ is the usual transpose of $A$ and  
we choose the $4\times 4$ matrix $K$ satisfying $K^2=-I$ to be
\begin{equation}
K={\scriptsize\left(
\begin{array}{cccc}
  0 & 1 & 0 & 0 \\
  -1 & 0 & 0 & 0 \\
   0 & 0 & 0 & 1 \\
   0 & 0 & -1 & 0
\end{array} \right)\, .}
\end{equation}
Any matrix $M$ from $\su(2,2|4)$ can then be decomposed into the sum 
$$
M=\underbrace{M^{(0)} + M^{(2)}}_{\mbox{even}}+ \underbrace{M^{(1)} + M^{(3)}}_{\mbox{odd}}\,,
$$
where every matrix $M^{(p)}$ is an eigenstate of $\Omega$
\begin{equation}
\label{Omeig}
\Omega(M^{(p)}) = i^p M^{(p)} \, .
\end{equation}
Explicitly the matrices $M^{(p)}$ are given by 
\bea
\label{M0}
M^{(0)} &=&{1\ov 4}\left( M + \Omega (M) + \Omega^2 (M)+ \Omega^3 (M)\right) =
{1\ov 2} \left(
\begin{array}{cc}
  A + K A^t K & 0 \\
  0 & D +K D^t K
\end{array} \right)\, ,~~~~~~~~~\\
\label{M2}
M^{(2)} &=& {1\ov 4}\left(M - \Omega (M) + \Omega^2 (M)- \Omega^3 (M)\right)  =
{1\ov 2}\left(
\begin{array}{cc}
  A - K A^t K & 0 \\
  0 & D -K D^t K
\end{array} \right)\, ,~~~~~~~~~~\\
\label{M1}
 M^{(1)} &=& {1\ov 4}\left(M -i \Omega (M) - \Omega^2 (M)+ i\Omega^3 (M) \right)=
{1\ov 2} \left(
\begin{array}{cc}
  0 & X + i K Y^t K  \\
  Y -i K X^t K & 0
\end{array} \right)\, ,~~~~~~~~~\\
\label{M3}
 M^{(3)} &=& {1\ov 4}\left(M +i \Omega (M) - \Omega^2 (M)-i\Omega^3 (M) \right)=
{1\ov 2} \left(
\begin{array}{cc}
  0 & X - i K Y^t K  \\
  Y +i K X^t K & 0
\end{array} \right)\, .~~~~~~~~~
\eea
It is not difficult to check by using these formulas 
that the matrices $M^{(0)}$ form 
the ${\so(4,1)}\times 
\so(5)$ subalgebra which we wish to mod out in the coset. We also see that the
matrices $M^{(1,3)}$ contain the odd matrices. Splitting $M$ into Grassmann
even and odd parts
$$
M = M_{{even}} + M_{{odd}}\,,\quad M_{{even}}=\left(
\begin{array}{cc}
  A & 0 \\
  0 & D
\end{array} \right)\,  ,\quad M_{{odd}}=\left(
\begin{array}{cc}
  0 & X \\
  Y & 0
\end{array} \right)\,  ,
$$
one can also rewrite the explicit expressions for $M^{(p)}$
in the following form 
\bea
\label{M01}
M^{(0)} &=& {1\ov 2} \left( M_{even} + K_8 M_{even}^t K_8 \right) 
\, ,\qquad
M^{(2)} = {1\ov 2} \left(M_{even} - K_8 M_{even}^t K_8 \right)\, ,~~~~~~~~~\\
\label{M11}
M^{(1)} &=& {1\ov 2} \left(M_{odd} + i \widetilde{K}_8 M_{odd}^t K_8\right) \, ,\qquad
M^{(3)} = {1\ov 2} \left(M_{odd} - i \widetilde{K}_8 M_{odd}^t K_8 \right)\, ,~~~~~~~~~
\eea
where $K$ and $\widetilde{K}$ are defined as
$$
K_8 = \left(
\begin{array}{cc}
  K & 0  \\
  0 & K
\end{array} \right)\, ,\qquad \widetilde{K}_8 = \left(
\begin{array}{cc}
  K & 0  \\
  0 & -K 
\end{array} \right)\, .
$$


The orthogonal complement $M^{(2)}$ of ${\rm so(4,1)}\times {\rm
so}(5)$ in $\su(2,2)\oplus \su(4)$ can be conveniently described
as follows. In appendix A we introduce the Dirac matrices for SO(5) $\gamma_s$,  
$s=1,2,3,4$ and $\gamma_5\equiv \Sigma$, which we all take to be hermitian.
These matrices obey the relations 
\bea 
K\gamma_s^t K=-\gamma_s\, ,  \qquad K\Sigma^t K=-\Sigma \, ,
\eea 
and, therefore, span the orthogonal complement to the Lie algebra $\so(5)$. 
The same matrices can be used to build the set of Dirac-matrices for
$\so(4,1)$, one takes $\{i\,\Sigma,\gamma_a\}$ with $a=1,2,3,4$.
Hence, we can represent any matrix $M^{(2)}$ from $\su(2,2|4)$ in the form
\bea
M^{(2)}= \left(
\begin{array}{cc}
 i\,t\,\Sigma + z_a \gamma_a& 0  \\
  0 & i\, \phi\,\Sigma + i\, y_s \gamma_s
\end{array} \right) 
+ i\, m_0 \,\mathbb{I}
\equiv x_M \S_M + \left(\begin{array}{cc}
 i\,t\,\Sigma & 0  \\
  0 & i\, \phi\,\Sigma 
\end{array} \right) + i\, m_0 \,\mathbb{I} \, , \nonumber
\eea
where $x_M=\{z_a, y_s\}$, $t$, $\phi$ and $m_0$ are real parameters of $M^{(2)}$, and the $8\times 8$ matrices
\bea\la{SM}
\S_M= \left\{\left(
\begin{array}{cc}
 \gamma_a& 0  \\
  0 & 0 
\end{array} \right)\,, \left(
\begin{array}{cc}
 0& 0  \\
  0 & i  \gamma_s
\end{array} \right)\right\}\,, \qquad
\Sigma_+= \left(\begin{matrix}
\S & 0 \cr 0 & \S\cr \end{matrix}\right )   \,, \qquad
\Sigma_-=  \left(\begin{matrix}
-\S & 0 \cr 0 & \S\cr \end{matrix}\right ) \, .
\eea
together with the U(1) generator $i\mathbb{I}$ form a basis of $M^{(2)}$ which shall
be of use in the sequel.

\subsection{Lagrangian and Coset Element}

Consider now a group element $g$ belonging to ${\rm PSU}(2,2|4)$
and construct the following current 
\bea 
\label{la} A=-g^{-1}{\rm
d}g=\underbrace{A^{(0)}+A^{(2)}}_{\rm even
}+\underbrace{A^{(1)}+A^{(3)}}_{\rm odd}\, ,
\eea 
where we also exhibited its $\mathbb{Z}_4$ decomposition. By
construction this current has zero-curvature.
The Lagrangian density for the superstring in $\AdS$ can then be written in
the form \cite{Metsaev:1998it, Roiban:2000yy} 
\bea 
\label{sLag} 
\L
=-\frac{1}{2}\sqrt{\lambda}\,\str \Big( \gamma^{\a\b}
A^{(2)}_{\a}A^{(2)}_{\b} +\kappa
\epsilon^{\a\beta}A^{(1)}_{\a}A^{(3)}_{\beta} \Big)\, , 
\eea 
which is the
sum of the kinetic and the Wess-Zumino terms and $\kappa$-symmetry
requires $\kappa=\pm 1$. Here we use the
convention $\epsilon^{01}\equiv\epsilon^{\tau\sigma}=1$ and $\gamma^{\a\b}= h^{\a\b}
\sqrt {-h}$ is the Weyl-invariant combination of the metric on the
string world-sheet with $\det\gamma=-1$.

\medskip

There are many different ways to parametrize the coset
element (\ref{ce}) related to each other by non-linear field
redefinitions. For example the authors of \cite{Alday:2005jm}, 
considered the parametrization for the coset element 
\bea\la{par1}
g=g(\theta)g(x) 
\eea 
where $\theta$ parametrizes the fermionic and $x$ the bosonic degrees of freedom.
This form is especially convenient to analyze the global symmetries of the Lagrangian
(\ref{sLag}) as the symmetries act linearly on the fermionic variables $\theta$. 
Due to this the fermions $\theta$ are charged 
under any U(1) subgroup of ${\rm PSU}(2,2|4)$ and in particular under the subgroups 
generated by shifts of the global time coordinate $t$ of $AdS_5$ and of an angle variable
$\p$ of ${\rm S}^5$. On the other hand, as was discussed in \cite{AFlc}, 
to impose the light-cone gauge it
is convenient to use fields neutral under these two U(1) subgroups. This requires 
to redefine the fermions by performing 
a similarity transformation of the matrix $g(\theta)$ \cite{AAFg}\footnote{In \cite{AAFg} 
a similarity transformation was used to make fermions neutral under all six U(1) subgroups of 
${\rm PSU}(2,2|4)$ that was necessary to apply TsT-transformations 
to derive the Green-Schwarz action on $\g$-deformed 
$\AdS$ backgrounds \cite{LM,F}.}. The result of this transformation is
equivalent to choosing a different coset element from the start, namely we
choose the element to be of the form
\begin{equation}
\label{param1}
g(\chi, x, t, \phi) = \Lambda(t, \phi) g(\chi) g(x) \,.
\end{equation}
Here $x^M=\{ z_a, y_{s}\}$, $a,s=1,\cdots ,4$, and the coordinates $t, z_a$ and 
$\phi, y_s$ parametrize $AdS_5$ and ${\rm S}^5$, respectively. The even
matrices $\Lambda(t, \phi)$ and $g(x)$ describe an
embedding of $\AdS$ into  $\mbox{SU(2,2)}\times \mbox{SU(4)}$ and 
$g(\chi)$ is a matrix which incorporates the 32
fermionic degrees of freedom.
The matrix $g(x)$ and the diagonal matrix $\Lambda$ are defined as
\begin{equation}
g(x)=\left(\begin{array}{rr}
g_a(z) ~&~ 0  \\
0 ~&~ g_s(y)  \\
\end{array}\right) \, ,
\end{equation}
\begin{eqnarray}
\label{our-para}
g_a(z) = {1\over \sqrt{1- {z^2 \over 4}}} (1 + {1\over 2} z_a \gamma_a) \,,\qquad
g_s(y) = {1\over \sqrt{1 + {y^2 \over 4}}} (1 + {i\over 2} y_s \gamma_s)\,,
\end{eqnarray}
and 
\begin{equation}
\Lambda(t, \phi)=\exp\Bigl [ \frac i 2\, t \,\left(\begin{matrix}
\S & 0 \cr 0 & 0\cr \end{matrix}\right ) +
 \frac i 2\, \phi \,\left(\begin{matrix}
 0& 0 \cr 0 & \S\cr \end{matrix}\right )\, \Bigr ] =
\exp\Bigl [ \frac i 2\, x_+\, \Sigma_+ + \frac i 2 \, x_-\, \Sigma_-
\, \Bigr ] \,,
\end{equation}
where we have introduced the light-cone coordinates $x_\pm$: $t=x_+-x_-$ and $\phi=x_++x_-$, and used the $8\times 8$ matrices $\Sigma_\pm$ of (\ref{SM}).

It is not difficult to check that the following important relations are valid
\bea\la{sg}
\S_\pm g^{-1}(x) = g(x)\S_\pm\,,
\eea
because $\S$ anticommutes with $\g_a$ and $g^{-1}(x) = g(-x)$.

Using the
parametrisation (\ref{our-para}) the metric on the coset becomes
\begin{eqnarray}
\label{metric}
ds^2 &=& -G_{tt}(z)
dt^2 + G_{\phi\phi} d\phi^2 + {1\over (1 - {z^2 \over 4})^2} dz_idz_j 
+ {1\over (1 + {y^2 \over 4})^2} dy_idy_j \, . \nonumber \\
G_{tt} &=& \left( {1 + {z^2 \over 4}  \over 1 - {z^2 \over 4}} \right)^2 \, 
\quad G_{\phi\phi} = \left( {1 - {y^2 \over 4}  \over 1 + {y^2 \over 4}} \right)^2\,,
\end{eqnarray} 
that shows explicitly that $t$ is the global time coordinate of ${\rm AdS}_5$, 
$\phi$ is an angle of ${\rm S}^5$, and $z_i$ and $y_i$ are the remaining
coordinates of ${\rm AdS}_5$ and ${\rm S}^5$, respectively.

Finally, we choose the matrix $g(\chi)$ to be of the form
\bea\la{chi}
g(\chi) = \chi + \sqrt{1+\chi^2}\,,
\eea
where the odd matrix $\chi$ is 
\begin{equation}
\label{fermb} \chi=\left(
\begin{array}{cc}
0 & \T \\
\Ts & 0 
\end{array}\right)\,,\quad \Ts = -\Td\S\,,\quad  \T =\left(
\begin{array}{cccc}
\theta_{11} & \theta_{12} & \theta_{13} & \theta_{14} \\
\theta_{21} & \theta_{22} & \theta_{23} & \theta_{24} \\
\theta_{31} & \theta_{32} & \theta_{33} & \theta_{34} \\
\theta_{41} & \theta_{42} & \theta_{43} & \theta_{44} 
\end{array}\right)
\, .
\end{equation}
Here $\theta_{ij}$ are  complex fermions, and $\Td$ is the hermitian conjugate 
of $\T$.
By construction the element $g$ and,
$g(\chi)$ in particular, belong to the supergroup
SU(2,2$|$4). Let us stress that the fermions and the bosonic coordinates 
$z_i$ and $y_i$ do not transform under the U(1) transformations 
generated by shifts of $t$ and $\phi$. The fields are charged 
under the four remaining U(1) subgroups of PSU(2,2$|$4). The charges 
are collected in Appendix A.

\section{Light-Cone Gauge }
In this section we introduce the first-order formalism for 
the \GS superstring in $\AdS$, and then, following \cite{AFlc}, 
impose the uniform light-cone  gauge and fix the kappa-symmetry. 
The uniform light-cone gauge generalizes
the standard phase-space light-cone gauge of \cite{Goddard:1973qh}
to a curved background, and differs from the one used in \cite{Metsaev:2000yu} 
by the choice of the light-cone coordinates and kappa-symmetry fixing. 
It belongs to the class of uniform gauges used to study the dynamics of spinning 
strings in $\AdS$ \cite{KT,AF}.
\medskip

\subsection{First-order Formalism}
The simplest way to impose a light-cone gauge is to introduce 
momenta canonically-conjugate to the coordinates $t$ and $\phi$ 
(or, equivalently, to the light-cone coordinates $x_\pm$).\footnote{This is the best way to 
impose any uniform gauge where a momentum is distributed uniformly along a string.}   
In the case of superstrings in $\AdS$ it is difficult to find the momenta because 
of a nontrivial interaction between bosonic and fermionic fields. A better way to proceed is to 
introduce a Lie-algebra valued auxiliary field $\bp$, 
and rewrite the superstring Lagrangian (\ref{sLag}) in the form
\begin{eqnarray}\nonumber
\L =  -\str \bigg( \bp \, A_0^{(2)} + 
\kappa{\sqrt{\lambda}\over 2}\epsilon^{\a\b}A_\a^{(1)} A_\b^{(3)}
 - {1\over 2 \sqrt{\lambda} \gamma^{00}} \left( \bp^2 + \lambda (A_1^{(2)})^2 \right)  
+ {\gamma^{01}\over \gamma^{00}} (\bp \, A_1^{(2)}) \bigg)\,. \\\label{Lang}
\end{eqnarray}
It is easy to see that if we solve the equations of motion for $\bp$
and substitute the solution back into (\ref{Lang}) one obtains (\ref{sLag}). 
The last two terms in (\ref{Lang}) yield the Virasoro constraints
\bea
\la{C1}
C_1 &=& \str\left( \bp^2 + \lambda (A_1^{(2)})^2 \right) =0\,,\\
\la{C2}
C_2 &=& \str\,\bp \, A_1^{(2)}=0\,,
\eea
which are to be solved after imposing the light-cone gauge and 
fixing the kappa-symmetry.

Without loss of generality we can assume that 
$\bp$ belongs to the subspace $M^{(2)}$ of $\su(2,2|4)$, as the other
components in the $\mathbb{Z}_4$ grading decouple. It therefore 
admits the following decomposition (compare (\ref{SM}) )
\begin{equation}
\label{bpexp}
\bp = \bp^{(2)} = {i\ov 4}\bp_+ \Sigma_+ +
 {i\over 4}\bp_- \Sigma_- + {1\over 2}\bp_M \Sigma_M 
 + \bp_0\, i\mathbb{I} \, .
\end{equation}
where $\Sigma^M$ are given by eq.(\ref{SM}).
It is obvious that since $A_\a^{(2)}$ belongs to the superalgebra $\su(2,2|4)$, $\str A_\a^{(2)} =0$, 
the variable $\bp_0$ does not contribute to the Lagrangian. 
The decomposition (\ref{bpexp}) secures the following identity 
\bea\la{strpba}
\str\, \bp \, A_\a^{(2)}= \str\, \bp^{(2)} \, A_\a = \str\, \bp\, A_\a\,.
\eea 

The fields $\bp_\pm$ are not the momenta $p_\pm$ canonically conjugate to 
$x_\mp$ but they may be 
expressed in terms of $p_\pm$. Before doing so let us impose the kappa-symmetry
gauge conditions, which simplify all expressions dramatically. 

\subsection{Fixing kappa-symmetry}
A key property of the \GS action is its invariance under the 
fermionic kappa-symmetry that halves the number of fermionic degrees of 
freedom. A kappa-symmetry gauge should be compatible with the bosonic
gauge imposed, and analyzing the kappa-symmetry transformations 
(which can be extracted from \cite{Metsaev:1998it, Roiban:2000yy}) for the \GS 
superstring 
action (\ref{Lang})  
one can show that in the case of the uniform
light-cone gauge kappa-symmetry can be fixed by choosing the 
fermion $\T$ of (\ref{fermb}) to be of the form 
\begin{equation}\T =\left(
\begin{array}{cccc}
0 & 0 & \theta_{13} & \theta_{14} \\
0 & 0 & \theta_{23} & \theta_{24} \\
\theta_{31} & \theta_{32} & 0 & 0 \\
\theta_{41} & \theta_{42} & 0 & 0 
\end{array}\right)
\, .
\end{equation}
It is not difficult to check that $\T$ of such a form anticommutes with $\S$ and 
therefore
the gauge-fixed $\chi$ satisfies the following important relations
\bea\la{Schi}
\S_+\chi = -\chi\S_+\,,\quad \S_-\chi = \chi\S_-\,.
\eea
In fact these relations may be considered as the defining relations for the 
kappa-symmetry gauge we have chosen and 
can be used instead of specifying the explicit form of $\chi$.
  
Taking into account that $g^{-1}(\chi)= g(-\chi)$ and these identities, one can easily show that
\begin{align}
g^{-1}(\chi)\S_+&=\S_+g(\chi)\quad \Rightarrow \quad g^{-1}(\chi)\S_+g(\chi)=
\S_+g(\chi)^2\,,\\
g^{-1}(\chi)\S_-&=\S_-g^{-1}(\chi)  \Rightarrow \quad 
g^{-1}(\chi)\S_-g(\chi)=\S_-\,.
\end{align}
Now it is straightforward to
use the coset parametrization (\ref{param1}) to 
compute the current (\ref{la})
\bea\nonumber
A &=& A_{even}+A_{odd}\\
\la{Aeven}
A _{even}&=&-g^{-1}(x)\Big[ {i\ov 2} dx_+ \S_+(1+ 2\chi^2) + {i\ov 2} dx_- \S_-
\Big]g(x)\\\nonumber
&~&-g^{-1}(x)\Big[ \sqrt{1+\chi^2}d\sqrt{1+\chi^2} - \chi d\chi +dg(x)g^{-1}(x)\Big]g(x)\,. \\
\la{Aodd}
A_{odd}&=&-g^{-1}(x)\Big[ i dx_+ \S_+ \chi\sqrt{1+\chi^2}+ \sqrt{1+\chi^2}d\chi-\chi d\sqrt{1+\chi^2}
\Big]g(x)\,. 
\eea
These formulas demonstrate explicitly the important advantage of the kappa-symmetry 
gauge we have 
imposed. The odd part of the current $A$ does not depend on the light-cone coordinate 
$x_-$! It also explains the drastic simplifications that will occur in the uniform 
light-cone gauge in comparison to the uniform gauge $t = \tau, \ p_\p=J$ used in 
previous works \cite{Callan, AF}. In the gauge 
$x_+ = \tau$  the odd part of $A$ depends only on the derivatives of the fermion $\chi$.
\subsection{Fixing the light-cone gauge}
After having fixed the kappa-symmetry we can now proceed to express $p_\pm$ in terms
of  $\bp_\pm$.
To this end, omitting the Virasoro constraints, we can rewrite the Lagrangian (\ref{Lang}) as follows
\bea\label{Lang2}
\L = p_+\dot{x}_- +{\bf p}_-\dot{x}_+ -\str \bigg( \bp A_{even}^\perp +
\kappa{\sqrt{\lambda}\over 2}\epsilon^{\a\b}A_\a^{(1)} A_\b^{(3)}\bigg)\,,
\eea
where 
\bea
A_{even}^\perp = -g^{-1}(x)\Big[ \sqrt{1+\chi^2}\pa_\tau\sqrt{1+\chi^2} - 
\chi \pa_\tau\chi +\pa_\tau g(x)g^{-1}(x)\Big]g(x)\,,
\eea
and
the momentum $p_+$ canonically conjugate to 
$x_-$ can be easily shown to be equal to
\bea\label{p+explicit}
p_+ = {i\ov 2} \text{Str} \left( \bp \Sigma_- g(x)^2 \right) 
= G_+ \bp_+ - G_- \bp_-\,,\qquad
G_\pm &=&{1\over 2}(G_{tt}^{1\over 2} \pm G_{\phi\phi}^{1\over 2}) \,.~~~~~~~~
\end{eqnarray}
The variable ${\bf p}_-$ is not equal to the momentum $p_-$ canonically conjugate to 
$x_+$. It differs from $p_-$ by a contribution coming from the Wess-Zumino term, 
and is defined as follows
\bea \la{bpm}
{\bf p}_-
={i\ov 2} \text{Str} \left( \bp \Sigma_+ g(x)(1+2\chi^2)g(x) \right) \,.
\eea
The uniform light-cone gauge is now imposed by setting \cite{AFlc}
\begin{eqnarray}
x_+ &=& \tau + {m\over 2} \sigma \,, \quad p_+ =P_+ = E + J = const\,,
\nonumber \\ x_{\pm} &=& {1\over 2} (\phi \pm t) \,, \quad  
p_+ = p_\phi - p_t \, \quad p_- = p_t + p_\phi \, ,
\end{eqnarray}
where the space-time energy $E$ and the angular momenta $J$ are integrals over 
$\s$ of the momenta $p_t$ and $p_\p$ conjugate to the global AdS time $t$ and the angle $\p$
\bea
E = -\int{d\s\ov 2\pi} p_t\,,\qquad J = \int{d\s\ov 2\pi} p_\p\,.
\eea 
The string winding number $m$ appears because $\p$ is an angle variable. 
In what follows we will be interested in the near plane wave limit, and, therefore, we 
set $m=0$. Let us stress again that the density $\H$ of the target 
space Hamiltonian is {\it not} equal to $-{\bf p}_-$. The Wess-Zumino term in (\ref{Lang2}) 
also contributes to $p_-$, and, therefore, to $\H$.

\subsection{PSU(2,2$|$4) charges}

The invariance of the \GS action under the PSU(2,2$|$4) group leads to 
the existence of conserved currents and charges. As was shown in
\cite{Alday:2005gi} the conserved currents can be written in terms of $A_\a$ as follows
\bea\la{Ccurr}
J^\a = \sqrt{\l} g(x,\theta)\left( \g^{\a\b} A_\b^{(2)} 
- {\kappa\ov 2} \epsilon^{\a\b}(A_\b^{(1)} - A_\b^{(3)})\right)g(x,\theta)^{-1}\,.
\eea
The conserved charges are then given by integrals over $\s$ of $J^\tau$
 \bea\la{Cchar}
Q = \int_0^{2\pi} {d\s\ov 2\pi}J^\tau \,.
\eea
For our purposes it is convenient to express the charges in terms of the momenta $\bp$. 
To this end we notice that they satisfy the following equations of motion
\bea\la{eqm}
\bp = \sqrt{\l} \g^{\tau\b}A_\b^{(2)} = \sqrt{\l} \g^{\tau\tau}\left( A_\tau^{(2)} 
+ {\g^{\tau\s}\ov \g^{\tau\tau}}A_\s^{(2)}\right)\,,
\eea
and therefore we can express $A_\tau^{(2)}$ in terms of $\bp$, and substitute it
into the expression (\ref{Cchar}) for $Q$. After a simple algebra we get
\bea
Q = \int_0^{2\pi} {d\s\ov 2\pi} g(x,\theta)\left(\bp 
- \sqrt{\l}{\kappa\ov 2} (A_\s^{(1)} - A_\s^{(3)})\right)g(x,\theta)^{-1}\,.
\eea
The formula can be written in a more explicit form if we take into account that
\bea
A_\s^{(1)} - A_\s^{(3)} = -i g(x)\widetilde{K}_8 F_\s^t K_8 g(x)^{-1}\,,
\eea
where 
\bea
F_\s = \sqrt{1+\chi^2}\pa_\s\chi - \chi\pa_\s\sqrt{1+\chi^2}\,,
\eea
is an odd component of the current $g^{-1}(\chi)\pa_\s g(\chi)$. Then, 
the $\psu(2,2|4)$ charges are
\bea\la{charges}
Q = \int_0^{2\pi} {d\s\ov 2\pi} \Lambda g(\chi) g(x)\left(\bp 
+i\sqrt{\l}{\kappa\ov 2}g(x)\widetilde{K}_8 F_\s^t K_8 g(x)^{-1} \right)
g(x)^{-1}g(\chi)^{-1}\Lambda^{-1}\,.~~~~~
\eea
The expression is very simple, and  it has the important property 
that $Q$ does not have an explicit dependence
on the world-sheet metric.

The combinations of components of the matrix $Q$ give charges corresponding 
to rotations, dilatations, supersymmetry and so on. To single out the charges one should
multiply $Q$ by a corresponding $8\times 8$ matrix, and take the supertrace
\bea
Q_{\cal M} = \str\, (Q{\cal M}) \,.
\eea
The diagonal and skew-diagonal  $4\times 4$ blocks of ${\cal M}$ single out  bosonic and 
fermionic charges of
$\psu(2,2|4)$, respectively. 

We divide all charges into two groups: kinematic and dynamic charges. Kinematic charges
are those that do not depend on $x_-$ and receive no corrections. 
The matrices ${\cal M}$ corresponding to kinematic charges are of the form
\bea\la{Mkin}
{\cal M}_{kin}=\left(
\begin{array}{cccc}
~a~ & ~0~ & ~0~ & ~g~ \\
~0~ & ~b~ & ~f~ & ~0~ \\
~0~ &~\tilde{f}~ & ~\tilde{a}~ & ~0~ \\
~\tilde{g}~ & ~0~ & ~0~ & ~\tilde{b}~ 
\end{array}\right)
\eea
where $a$, $b$, $\tilde{a}$ and $\tilde b$ are su(2) matrices.
This is because $\S_-$ commutes with any matrix of such a form, and 
therefore $x_-$ drops out of 
$Q_{\cal M}$. We also add to these charges the l.c. momentum $p_+$ that is 
expressed in terms of $Q$ as follows
\bea
p_+ = {i\ov 2}\str\, (Q\S_-)\,.
\eea
One can easily check that $p_+$ coincides with (\ref{p+explicit}).

The matrices ${\cal M}$ corresponding to dynamic charges are obviously of the form
\bea\la{Mdyn}
{\cal M}_{dyn}=\left(
\begin{array}{cccc}
~0~ & ~c~ & ~h~ & ~0~ \\
~d~ & ~0~ & ~0~ & ~k~ \\
~\tilde{h}~ &~0~ & ~0~ & ~\tilde{c}~ \\
~0~ & ~\tilde{k}~ & ~\tilde{d}~ & ~0~ 
\end{array}\right)
\, ,
\eea
and we also should add the l.c. momentum $p_- = -H_{l.c}$ that is expressed in terms of 
$Q$ as follows
\bea
p_- = {i\ov 2}\str\, (Q\S_+)\,.
\eea
It is not difficult to verify that  
$-p_-$ coincides with the Hamiltonian (\ref{H}).

A convenient basis of the matrices ${\cal M}$ is provided by the $SO(4)$ gamma matrices.
It is not difficult to check that ${\cal M}_{kin}$ are span by the following set of
29 matrices
\bea
{\cal M}_{kin} = \left\{\S_-\,, P_2^\pm\otimes {1\ov 2}[\G_a,\G_b]\,,
\s^\pm\otimes \G_a\,, \s^\pm\otimes \S\G_a\right\}\,,
\eea
and ${\cal M}_{dyn}$ are span by the following 33 matrices
\bea
{\cal M}_{dyn} = \left\{\S_+\,, P_2^\pm\otimes \G_a\,, P_2^\pm\otimes \S\G_a\,,
\s^\pm\otimes {1\ov 2}[\G_a,\G_b]\,, \s^\pm\otimes \S \,, \s^\pm\otimes I_4 \right\}\,.
\eea
The detailed structure of the conserved charges and their algebra will be discussed elsewhere.

\section{Gauge-fixed Lagrangian}
Now we are ready to find the light-cone gauge-fixed Lagrangian. 
This is a multistep procedure: First we solve equation  
(\ref{p+explicit})   for $\bp_+(P_+,\bp_-)$. Second
we solve the Virasoro constraint $C_2$ of equation (\ref{C2}) to find 
$x_-'$.  Finally 
we determine $\bp_-$ from the second Virasoro constraint $C_1$ eq.(\ref{C1}). 
Substituting all the solutions into the Lagrangian of equation 
(\ref{Lang2}), we end up with the total
gauge-fixed Lagrangian. This explicit derivation and definitions of all 
the quantities used to write it down may be found in Appendix B.

The upshot is a Lagrangian which can be written in the standard form as the difference
of a kinetic term and the Hamiltonian density
\bea
\label{Lgf}
&&\L_{gf} = \L_{kin} - \H
\,.~~~~~~~
\eea
The kinetic term $\L_{kin}$ depends on the time derivatives of the 
physical fields, and 
determines the Poisson structure of the theory. It can be cast in the form
\bea
\label{Lkin}
\L_{kin} &=& p_M\dot{x}_M - 
\frac{iP_+}{4}\str\left(\Sigma_+\chi\pa_\tau\chi\right)
+\frac{1}{2}g_N\bp_M\str\left(\left[\S_N,\Sigma_M\right] B_\tau\right)\\
\nonumber
&+& 
i\kappa{\sqrt{\lambda}\over 2}(G_+^2-G_-^2)\str\left( F_\tau\widetilde{K}_8
F_\s^t K_8\right)
-i\kappa{\sqrt{\lambda}\over 2} G_MG_N\str\left(\S_N F_\tau\S_M \widetilde{K}_8
F_\s^tK_8\right)
\,.~~~~~~~
\eea
where the functions $B_\alpha$ and $F_\alpha$ refer to the even and odd components
of $g^{-1}(\chi)\,\partial_\alpha  g(\chi)$ and are explicitly defined in the
appendix B.2 in eq.(\ref{BF}).
As one can see, the kinetic term is highly nontrivial and leads to a complicated Poisson structure 
(similar to the one derived in \cite{Alday:2005jm} for strings in $\su(1|1)$ subsector). To quantize 
the theory perturbatively, e.g. in the near plane-wave limit, 
we will need to redefine the fields  so that 
the kinetic term would acquire the conventional form
\bea\la{kincan}
\L_{kin}\rightarrow p_M\dot{x}_M - 
\frac{i}{2}\str\left(\Sigma_+\chi\pa_\tau\chi\right)\,,
\eea
and, therefore, the redefined fields would satisfy 
the canonical commutation relations. This will be done in the next section 
up to the quartic order in
the fields.

The density $\H$ of the Hamiltonian is given by the sum of $- {\bf p_-}$ and 
a contribution of the Wess-Zumino term
\bea
\label{H}
\H &=& - {\bf p_-} + \H_{WZ}\\\la{Hwz}
\H_{WZ}&=&\kappa{\sqrt{\lambda}\over 2}(G_+^2-G_-^2)\str\left(\S_+ 
\chi\sqrt{1+\chi^2}\widetilde{K}_8F_\s^t
K_8\right) 
\\
\nonumber
&~&+
\kappa{\sqrt{\lambda}\over 2}G_MG_N\str\left(\S_+\S_N\chi\sqrt{1+\chi^2} \S_M \widetilde{K}_8
F_\s^t K_8\right)
\,.~~~~~~~
\eea
Here the explicit expression for ${\bf p_-}$ is given by (\ref{pm1B}), and we also 
should use 
the formulas (\ref{l15}), (\ref{l13}) and (\ref{l27}) 
to express everything in terms 
of physical fields. 

Let us stress that in this way we find the gauge-fixed Lagrangian as an {\it exact} 
function of the light-cone momentum $P_+$, and the string tension $\sqrt\l$. 
Then it is straightforward  to consider
various expansions of the Lagrangian, in particular, 
in the next sections we will consider a near plane-wave expansion, with
$P_+\to\infty$ with $\lambda/P_+^2$ fixed.

\section{Near plane-wave expansion }
In this section we discuss the near plane-wave expansion of the 
gauge-fixed Lagrangian (\ref{Lgf}), which amounts to a large $P_+$ limit
with the effective coupling $\tl = {4\l\ov P_+^2}$ fixed. 
For this we need to redefine the fields
to reduce the kinetic term (\ref{Lkin}) to the canonical form (\ref{kincan}) up to 
terms of sixth order in fields, and 
expand the density of the Hamiltonian to the quartic order.
The resulting quartic Hamiltonian may be used to compute 
the $1/J$ correction to the space-time energy $E$ of an arbitrary string state.

\subsection{Field redefinition}
It is straightforward to find the necessary field redefinition 
to remove all quartic terms in the kinetic term of the Lagrangian.
To this end we notice that the kinetic term can be written in the following form
\bea
\label{Lkin2}
\L_{kin} &=& p_M\dot{x}_M - 
\frac{iP_+}{4}\str\left(\Sigma_+\chi\dot\chi\right)
+\frac{iP_+}{2}\str\left(\Sigma_+\Phi(p,x,\chi)\dot\chi\right)
\,.~~~~~~~
\eea
where $\Phi$ is a function of cubic order in physical fields and
has the same structure 
as $\chi$, i.~e.~it satisfies the defining relations (\ref{Schi}) for
our kappa-symmetry gauge. Then it is clear that
the last term can be removed by the following redefinition of $\chi$
\bea\la{chired}
\chi\rightarrow \chi + \Phi(p,x,\chi)\,.
\eea
This redefinition casts the kinetic term (\ref{Lkin2}) into the 
form (up to a total derivative)
\bea
\label{Lkin3}
&&\L_{kin} = p_M\dot{x}_M - 
\frac{iP_+}{4}\str\left(\Sigma_+\chi\dot\chi\right)\\\nonumber
&&+\frac{iP_+}{2}\str\left(\Sigma_+\left(\Phi(p,x,\chi+\Phi) - \Phi(p,x,\chi)\right)\dot\chi\right)
+\frac{iP_+}{4}\str\left(\Sigma_+\Phi(p,x,\chi)\dot\Phi(p,x,\chi)\right)
.~~~~~~~
\eea
Since $\Phi$ is at least of the cubic order in the fields, 
the terms on the second line of (\ref{Lkin3}) are at least of the sixth order. 
These terms can be also removed by a similar field redefinition. However, this time one would need to 
redefine not only the fermions but the bosonic coordinates $x_M$ and $p_M$, too. 
For our purposes here it is sufficient to perform only the simplest redefinition (\ref{chired}), 
and just drop the terms on the second line of (\ref{Lkin3}). This reduces the kinetic term to 
the canonical quadratic form which can be written in a very explicit form as 
 \bea
\label{Lkin4}
\L_{kin} = p_M\dot{x}_M - 
\frac{iP_+}{4}\str\left(\Sigma_+\chi\dot\chi\right) &=& p_M\dot{x}_M +
\frac{iP_+}{4}\tr\left( \etad\dot{\eta} +\td\dot{\theta}\right)\\\nonumber
&=&p_M\dot{x}_M +
\frac{iP_+}{2}\left( \etad_a\dot{\eta}_a +\td_a\dot{\theta}_a\right)
\,.~~~~~~~
\eea  
Here we used the following decomposition of the fermions (see appendix A)
\bea\la{chiT}
\chi = \s_+\otimes \Theta + \s_-\otimes \Theta_*\, ,\quad 
\Theta &=& \PP_+\eta + \PP_-\td\, ,\quad \eta= \eta_a\G_a \, ,\quad \theta= \theta_a\G_a\,.
\eea
Rescaling the fermions 
\bea\la{ferresc}
\chi\rightarrow \sqrt{{2\ov P_+}}\chi\, ,\quad \eta\rightarrow \sqrt{{2\ov P_+}}\eta 
\, ,\quad \theta\rightarrow \sqrt{{2\ov P_+}}\theta \,, 
\eea
brings the kinetic term into the canonical form 
\bea
\label{Lkin5}
\L_{kin} = p_M\dot{x}_M \,+\,
i\,\etad_a\dot{\eta}_a \,+\,i\,\td_a\dot{\theta}_a
\,,~~~~~~~
\eea  
which shows that $(p,x)$, $(\etad,\eta)$ and $(\td,\theta)$ are canonically conjugate 
pairs.

\medskip

Before discussing the expansion of the Hamiltonian let us mention an important and 
nice property of the redefinition (\ref{chired}). One can check that up to  
sixth order in fields, the formula (\ref{l13}) for
$x_-'$ takes the form
\bea
\la{xmpred}
x_-' = - {1\ov P_+}\left( p_Mx_M' - 
\frac{i}{4}P_+\,\str\left(\Sigma_+\chi\chi'\right) +\pa_\s f(p,x,\chi)\right)
\, ,
\eea 
where $f(p,x,\chi)$ is a function of the momenta and coordinates. Thus, we see that integrating 
(\ref{xmpred}) over $\s$ we get the usual ``flat space'' level-matching condition
\bea
\la{lm}
{\cal V}=\int_0^{2\pi}\left( p_Mx_M' - 
\frac{i}{4}P_+\,\str\left(\Sigma_+\chi\chi'\right)
\right) = 0\,,
\eea
which in terms of the rescaled fermions (\ref{ferresc}) takes the form
\bea
\la{lmr}
{\cal V}=\int_0^{2\pi}\left( p_Mx_M' + i \etad_a\eta'_a \,+\,i\,\td_a\theta'_a
\right) = 0\,.
\eea

\subsection{Hamiltonian}
Here we derive the density $\H$ of the Hamiltonian up to the fourth order in fields. 
To this end we expand (\ref{H}), and take into account 
the fermion redefinition (\ref{chired}). The fermion shift produces additional quartic terms in the Hamiltonian coming only from the shift of the quadratic part at
this order. 

\medskip

The density of the complete quadratic Hamiltonian can be easily 
found by using formulas from appendix B
\bea
\H_2 = {1\ov P_+} p_M^2 + {P_+\ov 4}x_M^2 +{\l\ov P_+}x_M'^2  + {\kappa\ov 2}\sqrt{\l}
\str \left(\S_+ \chi\widetilde{K}_8\chi'^t K_8\right) + {P_+\ov 4}\str\,\chi^2\, .\la{Hquadr}
\eea
We see from this equation and the kinetic term (\ref{Lkin4}) 
that in order to have a canonical Poisson structure and a standard 
quadratic Hamiltonian  of the form ${1\ov 2} p_M^2 + {1\ov 2}x_M^2$
we should make the following rescaling of the fields
\bea
\la{resc}
p_M \to \sqrt{{P_+\ov 2}}p_M\,,\quad x_M \to\sqrt{{2\ov P_+}}x_M\,,\quad \chi\to\sqrt{{2\ov P_+}}\chi\,.
\eea
Then the density of the quadratic Hamiltonian takes the form
\bea\la{Hquadr2}
\H_2 = {1\ov 2} p_M^2 + {1\ov 2}x_M^2 +{\tl\ov 2}x_M'^2  + {\kappa\ov 2}\sqrt{\tl}
\str \left(\S_+ \chi\widetilde{K}_8\chi'^t K_8\right) + {1\ov 2}\str\,\chi^2\, ,
\eea
where
\bea\la{tla}
\tl = {4\l\ov P_+^2}\, 
\eea
is the effective coupling constant which is kept finite in the plane-wave 
limit 
$P_+\to\infty$ or equivalently $\l\to\infty$. 
This limit is the light-cone gauge equivalent of the usual $J\to\infty$ BMN limit.
Note that the coupling $\tl$ is not equal to the effective coupling 
$\l' = {\l\ov J^2}$ but reduces to it only in the strict $J\to\infty$ limit.

In terms of the rescaled bosons and fermions $\eta$ and $\theta$ 
$\H_2$ takes the form
\bea\la{Hquadr3}
\H_2 = {p_M^2\ov 2}  + {x_M^2\ov 2} +{\tl\ov 2}x_M'^2  + 
{1\ov 2}\tr\left( \etad\eta +\td\theta 
+{\kappa\sqrt{\tl}\over 2}\left(  \eta\eta' +\theta\theta'-\etad\etapd -\td\tpd\right) \right)\, .
~~~~~~
\eea

\medskip

The quartic Hamiltonian is also straightforwardly derived. The details of the 
computation can be found in appendices C and D. In terms of the rescaled fields 
(\ref{resc}) it takes the form
\bea\nonumber
\H_4&=&{1\ov 2P_+}\Big[\, 2\tl \left(y'^2z^2-z'^2y^2+
z'^2 z^2 - y'^2 y^2\right)\\\nonumber
 &-&\tl\str\left({1\ov 2}\chi\chi'\chi\chi' + \chi^2\chi'^2+
{1\ov 4}(\chi\chi'-\chi' \chi)K_8(\chi\chi'-\chi' \chi)^tK_8
 +
\chi\widetilde{K}_8\chi'^t K_8 \chi\widetilde{K}_8\chi'^t K_8 \right) \\\nonumber
&+& \tl\str\left( (z^2-y^2)\chi'\chi'
+{1\ov 2} x_M' x_N \left[\S_M,\S_N\right](\chi\chi'- \chi'\chi)
-2 x_Mx_N\S_M\chi'\S_N\chi'\right) 
\\\la{H41}
&+&{i\kappa\sqrt{\tl}\ov 4}(x_Np_M)'\str\left( 
\left[\S_N,\Sigma_M\right]\left(\widetilde{K}_8\chi^t K_8\chi - \chi\widetilde{K}_8\chi^t K_8\right) 
\right) \Big]  \, .
\eea 
One can easily see that the uniform light-cone gauge quartic Hamiltonian (\ref{H41})  
is considerably simpler than the quartic Hamiltonian obtained by
Calan et.~al.\cite{Callan}, though still rather  involved. 
An important property of the Hamiltonian is that it vanishes in the
point particle limit, when all fields do not depend on $\s$. 
One can show that the same property is also valid for the sixth-order Hamiltonian. 
This is in accord with the observation that already the quadratic particle 
Hamiltonian reproduces the spectrum of type IIB supergravity on $\ads$.

It is convenient to express the Hamiltonian in terms of the (rescaled) 
complex bosonic fields 
$Z_a$, $Y_a$ (see also appendix A) 
\bea
Z_1 &=& z_2 + iz_1\, ,\quad Z_2 = z_4 + iz_3\, ,\quad 
Z_{4} = z_2 - iz_1\, ,\quad Z_{3} = z_4 - iz_3\, ,
\\\nonumber
Y_1 &=& y_2 + iy_1\, ,\quad Y_2 = y_4 + iy_3\, ;\quad 
Y_{4} = y_2 - iy_1\, ,\quad Y_{3} = y_4 - iy_3\,.
\eea
and their canonical momenta (associated to
$z_a$ or $y_a$)
\bea
P_1 = {1\over 2} (p_2 + i p_1)\, , \quad P_{4} 
= {1\over 2} (p_2 -  i p_1) \, , \quad
P_2 = {1\over 2} (p_4 + i p_3) \, , \quad  P_{3} 
= {1\over 2} (p_4 -  i p_3) \, ,\nonumber
\eea
with 
\begin{equation}
Z^\dagger_a=Z_{5-a}\, , \quad Y^\dagger_a=Y_{5-a}\, ,\quad P^{z\,\dagger}_a=P^z_{5-a}\, , 
\quad P^{y\,\dagger}_a=P^y_{5-a}\, .
\end{equation}
The convention is chosen such 
that $p_M\, x_M = P^z_{5-a}\, Z_a +P^y_{5-a}\, Y_a $. Then the kinetic term (\ref{Lkin5}) takes the form
\bea
\label{Lkin6}
\L_{kin} = P_{5-a}^z\dot{Z}_a + P_{5-a}^y\dot{Y}_a + 
i\etad_a\dot{\eta}_a +i\td_a\dot{\theta}_a
\qquad \mbox{with}\quad a=1,2,3,4\,.~~~~~~~
\eea  
In terms of the complex fields the quadratic Hamiltonian acquires the form
\bea\nonumber
\H_2 &=& P_{5-a}^zP_a^z +  P_{5-a}^yP_a^y +
{1\ov 4}\left(Z_{5-a}Z_a + Y_{5-a}Y_a\right) 
+{\tl\ov 4}\left(Z_{5-a}'Z_a' + Y_{5-a}'Y_a'\right)  \\&+& 
{1\ov 2}\tr\left( \etad\eta +\td\theta 
+{\kappa\sqrt{\tl}\over 2}\left(  \eta\eta' +\theta\theta'-\etad\etapd -\td\tpd\right) \right)
\, ,
\la{Hquadr4}
\eea
and the quartic Hamiltonian is given by the following sum 
\bea
\H_4 = \H_{bb} + \H_{bf}+\H_{ff}\, ,
\eea
where
\bea\la{Hbb}
\H_{bb} 
= {\tl\ov 4P_+}
\left(Y_{5-a}'Y_a'Z_{5-b}Z_b - Y_{5-a}Y_aZ_{5-b}'Z_b'  
+ Z_{5-a}'Z_a'Z_{5-b}Z_b - Y_{5-a}'Y_a'Y_{5-b}Y_b\right),~~~~~~
\eea
\bea\la{Hbf2}
\H_{bf} &=& {1\ov 2P_+}\tr\Big[
{\tl\ov 2} (Z_{5-a}Z_a-Y_{5-a}Y_a)\left(\etapd\eta' +\tpd\theta'\right)\\\nonumber
&~&~~~~~ -{\tl\ov 2}\,
Z_m'Z_n\left[\G_m,\G_n\right] \left(\PP_+(\eta\etapd - \eta'\etad)- 
\PP_- (\td\theta' -\tpd\theta)\right)\\\nonumber
&~&~~~~~ + {\tl\ov 2}\,
Y_m'Y_n\left[\G_m,\G_n\right]\left(- \PP_-(\etad\eta' -\etapd\eta) 
+ \PP_+(\theta\tpd -\theta'\td)\right)
\\\nonumber
&~&~~~~~ - {i\kappa\ov 2}\sqrt{\tl}\,(Z_n P^z_m)'\left[\G_n,\G_m\right]\left(
\PP_+( \etad\etad +\eta\eta) +
\PP_-( \td\td +\theta\theta)\right)
\\\nonumber
&~&~~~~~
+{i\kappa\ov 2}\sqrt{\tl}\,(Y_n P^y_m)'\left[\G_n,\G_m\right]\left(\PP_-( \etad\etad +\eta\eta) 
+\PP_+( \td\td +\theta\theta) \right)\\\nonumber
&~&~~~~~ +4i\tl Z_m Y_n\left(- \PP_- \G_m\eta'\G_n\theta'
+ \PP_+ \G_m\tpd\G_n\etapd \right)\Big]\, ,~~~~~~~
\eea
The quartic fermionic term can be written in the form
\bea\la{hff}
\H_{ff} = \H_{ff}(\eta) -  \H_{ff}(\theta)\, ,
\eea
where $H_{ff}(\eta)$ takes the following amazingly simple form
\bea\la{hffeta2}
\H_{ff}(\eta)= -{\tl\ov 4P_+}\tr\,
\S\left(\etapd\eta\etapd\eta + \etad\eta'\etad\eta' + 
\etapd\etad\etapd\etad + \eta'\eta\eta'\eta
\right)\, .
\eea

\section{Quantization}

We now turn to the perturbative quantization of the light-cone
$\AdS$ superstring in the near plane wave limit. Due to the fermionic
field redefinitions performed above up to sixth-order the kinetic
Lagrangian is of canonical form. Promoting all fields to operators, we 
read off from (\ref{Lkin6}) the (anti)commutation relations
\begin{equation}
[Z_a,P^z_{5-b}] = i\, \delta_{ab} \qquad [Y_a,P^y_{5-b}] = i\,\delta_{ab} \qquad
\{\eta_a,\eta_b^\dagger\} = \delta_{ab} \qquad \{\theta_a,\theta^\dagger_b\}=\delta_{ab}
\, .
\end{equation}
We now need to establish a mode decomposition of the bosonic and fermionic
fields which renders the quadratic piece of the Hamiltonian (\ref{Hquadr4}) 
in a diagonal form. This we will be done for the bosonic and fermionic
sector in the following.

\subsection{Representation for bosons}

The bosonic part of the quadratic Hamiltonian (\ref{Hquadr4}) has the form
\begin{equation} 
H_{bos}^{(0)}=  P_{5-a}^zP_a^z +  P_{5-a}^yP_a^y +
{1\ov 4}\left(Z_{5-a}Z_a + Y_{5-a}Y_a\right) 
+{\tl\ov 4}\left(Z_{5-a}'Z_a' + Y_{5-a}'Y_a'\right) \, .
\end{equation}
We shall choose the following mode decompositions for $Z_a$ and $P^z_a$ 
\begin{align}
\nonumber
Z_a(\tau,\sigma) &= \sum_n e^{in\s} Z_{a,n}(\tau)\,,
\qquad &Z_{a,n}^\dagger(\tau,\sigma) = Z_{5-a,-n}(\tau.\sigma) \,;\\ \nonumber
P_a^z(\tau,\sigma) &= \sum_n e^{in\s} P_{a,n}^z(\tau) \,,\qquad 
&P_{a,n}^{z~\dagger}(\tau,\sigma) = P_{5-a,-n}^z(\tau,\sigma) 
 \, ,\\
P_{a,n}^z &=  {\sqrt{\om_n}\ov 2}\left( \b_{a,n}^++\b_{5-a,-n}^-\right)\,,\quad
&Z_{a,n} =  {1\ov i\sqrt{\om_n}}\left( \b_{a,n}^+-\b_{5-a,-n}^-\right)\,,
\end{align}
and similarly for $Y_a$ and $P^y_a$ 
\begin{align}
\nonumber
Y_a(\tau,\sigma) &= \sum_n e^{in\s} Y_{a,n}(\tau) 
\,,\qquad  &Y_{a,n}^\dagger(\tau,\sigma) = Y_{5-a,-n}(\tau,\sigma) \,;\\ \nonumber
& P_a^y(\tau,\sigma) = \sum_n e^{in\s} P_{a,n}^y(\tau)
\,,\quad & P_{a,n}^{y~\dagger}(\tau,\sigma) = P_{5-a,-n}^y(\tau,\sigma) 
 \, ,\\
P_{a,n}^y& =  {\sqrt{\om_n}\ov 2}\left( \a_{a,n}^++\a_{5-a,-n}^-\right)\,,\quad
&Y_{a,n} =  {1\ov i\sqrt{\om_n}}\left( \a_{a,n}^+-\a_{5-a,-n}^-\right)\,,
\end{align}
where the frequency $\om_n$ is defined as
\bea\la{omn}
\om_n = \sqrt{1+\tl\, n^2}\, .
\eea
Then in terms of the creation and annihilation operators the 
quadratic bosonic Lagrangian, steming of the kinetic piece (\ref{Lkin6})
and Hamiltonian (\ref{Hquadr}), takes the form
\bea
L_{bos}^{(0)}  &=& i\sum_{a,n}\left( \a_{a,n}^+\dot{\a}_{a,n}^- 
+\b_{a,n}^+\dot{\b}_{a,n}^-\right) - \sum_{a,n}\om_n\left( \a_{a,n}^+\a_{a,n}^- 
+\b_{a,n}^+\b_{a,n}^- \right)\,.
\eea
This shows that in quantum theory the only nontrivial commutators are simply
\bea
[ \a_{a,n}^-, \a_{a,n}^+] =1\,,\quad [ \b_{a,n}^-, \b_{a,n}^+] =1\,,
\eea
and we have the standard quadratic Hamiltonian.

\subsection{Representation for fermions}
For the quadratic fermionic sector we have the Hamiltonian from (\ref{Hquadr4})
\bea
H_{ferm}^{(0)}&=&{1\ov 2}\tr\left( \etad\eta +\td\theta 
-{\kappa\over 2}\sqrt{\tl}\left( - \eta\eta' -\theta\theta'+\etad\etapd +\td\tpd\right) \right)\,.
\eea
Our mode decomposition for fermions follows a similar construction 
found in \cite{Alday:2005jm} and reads
\begin{align}
\eta(\tau,\sigma) &= \sum_n e^{in\s} \eta_n(\tau)\,,\qquad   &\etad(\tau,\sigma) = \sum_n e^{-in\s} \etad_n(\tau)\, ,
\nonumber \\
\theta(\tau,\sigma) &= \sum_n e^{in\s} \theta_n(\tau)\,,\quad  & 
\td(\tau,\sigma) = \sum_n e^{-in\s} \td_n(\tau)\, \\
\eta_n &=  f_n\eta_{-n}^- + ig_n \eta_{n}^+\,,\quad & \etad_n = f_n\eta_{-n}^+ - ig_n \eta_{n}^-\,.
\nonumber \\
\theta_n &=  f_n\theta_{-n}^- + ig_n \theta_{n}^+\,,\quad &
\td_n = f_n\theta_{-n}^+ - ig_n \theta_{n}^-\,.
\end{align}
Here we have introduced the quantities
\be
f_n = \sqrt{{1\ov 2}\left(1+{1\ov \om_n}\right)}\,,\qquad 
g_n = {\kappa\sqrt{\tl} n\ov 1+\om_n}f_n \,.
\ee
note that $g_n^2 ={1\ov 2}\left(1-{1\ov \om_n}\right)$. 
In terms of the creation and annihilation operators the quadratic fermion Lagrangian then 
indeed takes the diagonalized form
\bea
L_{ferm}^{(0)}  &=& {i\ov 2}\tr\sum_n\left( \eta_n^+\dot{\eta}_n^- 
+\theta_{n}^+\dot{\theta}_n^-\right) - {1\ov 2}\tr\sum_n\om_n\left( \eta_n^+\eta_n^- 
+\theta_{n}^+\theta_n^-\right)\,.
\eea
If furthermore we use the decomposition of the matrices $\eta$ and $\theta$ in terms
of the Dirac matrices of appendix A as
\bea
\eta_n^- = \eta_{a,n}^-\G_{5-a}\,,\quad \eta_n^+ = \eta_{a,n}^+\G_{a}\,,\quad
\theta_n^- = \theta_{a,n}^-\G_{5-a}\,,\quad \theta_n^+ = \theta_{a,n}^+\G_{a}\,,
\eea
and the identity
$
\tr \G_a\G_{5-b}= 2\delta_{ab}\,,
$
then the quadratic Lagrangian may be rewritten as
\bea
L_{ferm}^{(0)}  &=& i\sum_n\left( \eta_{a,n}^+\dot{\eta}_{a,n}^- 
+\theta_{a,n}^+\dot{\theta}_{a,n}^-\right) -\sum_n\om_{n}\left( \eta_{a,n}^+\eta_{a,n}^- 
+\theta_{a,n}^+\theta_{a,n}^-\right)\,.
\eea
This shows that in quantum theory the only nontrivial anti-commutators between fermionic
mode operators are
\bea
\{ \eta_{a,n}^-, \eta_{a,n}^+\} =1\,,\quad \{ \theta_{a,n}^-, \theta_{a,n}^+\} =1\,,
\eea
and we have a standard diagonal quadratic Hamiltonian in the fermionic sector as well.

Note that we will take the quartic Hamiltonian $H_{bb} + H_{bf} + H_{ff}$ to be normal-ordered 
with respect to these bosonic and fermionic oscillator modes.
 
\subsection{Generic string state}
The generic eigenstate of the quadratic Hamiltonian can now be written in the form
\bea\la{gstate}
|\Psi \rangle 
=  \prod_{c=1}^4 |\theta_c, M^c_\theta \rangle \otimes
\prod_{c=1}^4 |\eta_c, M^c_\eta \rangle  
\otimes \prod_{c=1}^4|Z_c, M^c_z \rangle \otimes\prod_{c=1}^4
 |Y_c, M^c_y \rangle\,,
\eea 
where we assume the products to be in decreasing order 
$
\prod_{c=1}^4 f_c \equiv f_4f_3f_2f_1\,,
$
and take
\bea
|\theta_c, M^c_\theta \rangle \equiv \theta_{c}^+(n_{M^c_\theta})  
\theta_{c}^+(n_{{}_{M^c_\theta-1}}) \cdots  \theta_{c}^+(n_{{2}})   
\theta_{c}^+(n_{{1}})|0 \rangle\,,
\eea
\bea
|\eta_c, M^c_\eta \rangle \equiv \eta_{c}^+(m_{{}_{M^c_\eta}})  
\eta_{c}^+(m_{{}_{M^c_\eta-1}}) \cdots  \eta_{c}^+(m_{{2}})   
\eta_{c}^+(m_{{1}})|0 \rangle\,,
\eea
\bea
|Z_c, M^c_z \rangle \equiv \b_{c}^+(l_{{}_{M^c_z}})  
\b_{c}^+(l_{{}_{M^c_z-1}}) \cdots  \b_{c}^+(l_{{2}})   
\b_{c}^+(l_{{1}})|0 \rangle\,,
\eea
\bea
|Y_c, M^c_y \rangle \equiv \a_{c}^+(k_{{}_{M^c_y}})  
\a_{c}^+(k_{{}_{M^c_y-1}}) \cdots  \a_{c}^+(k_{{2}})   
\a_{c}^+(k_{{1}})|0 \rangle\,.
\eea
In the above we have used the notation
$
\eta_{c}^+(m)\equiv \eta_{c,m}^+\, \mbox{and so on}\,,
$
and we assumed that the mode numbers form increasing sequences, i.~e.~for fermions
$$
n_1 < n_2 <\cdots < n_{{}_{M^c_\theta-1}} < n_{{}_{M^c_\theta}}\,,\quad 
m_1 < m_2 <\cdots < m_{{}_{M^c_\eta-1}} < m_{{}_{M^c_\eta}}
$$
and for bosons
$$
l_1 \le l_2 \le\cdots \le l_{{}_{M^c_z-1}} \le l_{{}_{M^c_z}}\,,\quad 
k_1 \le k_2 \le\cdots \le k_{{}_{M^c_y-1}} \le k_{{}_{M^c_y}}\,.
$$
The energy of this state is 
\bea
H_2|\Psi \rangle = E|\Psi \rangle\,,\quad E= \sum_{\small\rm\ mode\ numbers} 
\om_{\rm mode\ number}\,.
\eea
The string states must also satisfy the level-matching condition (\ref{lm})
that in terms of the creation and annihilation operators takes the form
\bea
\la{lm3}
{\cal V} =\sum_{a,n}\, n\left( \a_{a,n}^+ \a_{a,n}^-  
+ \b_{a,n}^+ \b_{a,n}^- + \eta_{a,n}^+ \eta_{a,n}^-
+ \theta_{a,n}^+ \theta_{a,n}^-\right)\,.~~~~~~
\eea 
It just says that the sum of all mode numbers vanishes
\bea
{\cal V}|\Psi \rangle = 0\ \Rightarrow \sum_{\rm all\ mode\ numbers} (
\mbox{mode\ number}) = 0\, .
\eea

\section{Sectors and $1/J$ correction}
It is known that in \N SYM there are sectors of operators closed under the action of 
the dilatation operator, see \cite{revs} and references therein. 
In this section we explain how string states dual to operators 
from $\su(2)$, $\sl(2)$, $\su(1|1)$, $\su(1|2)$ and $\su(2|3)$ sectors 
can be constructed starting from 
corresponding eigenstates of the quadratic Hamiltonian, and compute $1/J$ corrections 
to energies of the states in  the $\su(2)$, $\sl(2)$, $\su(1|1)$ sectors.

\subsection{$\su(2)$ sector}
The $\su(2)$ sector of \N SYM consists of operators of the form
\bea\la{su2oper}
O_{su(2)} = \Tr \left( Z^J X^M + \mbox{permutations} \right)\,,
\eea
where $Z$ and $X$ are the two complex scalars carrying unit charges under 
the two U(1) subgroups of $SU(4)$ that in the string picture correspond to 
the U(1) generating 
shifts of the angle $\phi$ of ${\rm S}^5$ and the U(1) generated by $\Phi_1^S$, 
respectively, 
see appendix A. The operators, correspondingly, carry $J$ and $M$ units of 
charges.
They
are highest-weight which means
that they have minimal conformal dimensions among all the 
operators with the given charges.

Dual string states can be easily identified in the BMN limit $P_+\to\infty$, $\tl$ fixed. 
First of all the charge $J$ is assigned to 
the light-cone vacuum and {\it no} creation and annihilation 
operator carries charges under 
this U(1). Then from the tables of charges in appendix A we see that the string state carrying
$M$ units of charge under the second U(1) and having the minimal energy is obtained by 
acting on the vacuum by $M$ creation operators $\a_{1,n}^+$. Therefore, in the BMN limit 
the string states dual to operators from the $\su(2)$ sector are the states
\bea\la{su2state}
|\Psi_{su(2)} \rangle = \a_{1,n_M}^+ \a_{1,n_{M-1}}^+ \cdots \a_{1,n_1}^+  
|0 \rangle\,,
\eea
which are eigenstates of the quadratic Hamiltonian with 
the energy 
\bea\la{enersu2state}
E_0= \sum_{k=1}^M \om_{n_k}\,,
\eea
and satisfy the level-matching condition
\bea\nonumber
\sum_{k=1}^M n_{k} = 0\,.
\eea
For generic values of the mode numbers $n_k$ there is no nontrivial 
degeneracy in the spectrum,\footnote{Given a state with mode numbers $\{n_k\}$, the state with 
the mode numbers $\{-n_k\}$ has the same energy. One can easily see that the states do not 
mix with each other. One can also have a situation when mode numbers are divided into several 
groups, each group satisfying the level-matching condition. 
Then changing the signs of the mode numbers in any of the groups leads to a state with 
the same energy. One can show that these states do not mix at least at the $1/P_+$ order.} and 
the leading $1/P_+$ correction to the energy of the string state can be found just by computing
the average of the quartic Hamiltonian $H_{bb}$ in the state (\ref{su2state}). 
The computation is very simple because there is only one term in $H_{bb}$ contributing to 
the average. Explicitly we find
\bea
\nonumber
&&\langle \Psi_{su(2)}|H_{bb}|\Psi_{su(2)} \rangle = 
-{\tl\ov 4P_+}\int_0^{2\pi}{d\s\ov 2\pi}
\langle \Psi_{su(2)}|Y_{a}'Y_{5-a}'Y_{b}Y_{5-b}|\Psi_{su(2)} \rangle~~~~~\\\nonumber
&&~~~~~~~~~~~~={\tl\ov 4P_+}\sum_{n+m+k+l=0}{nm\ov\sqrt{\om_n\om_m\om_k\om_l}}
\langle \Psi_{su(2)}|4\a_{a,n}^+\a_{b,k}^+\a_{a,-m}^- \a_{b,-l}^-|\Psi_{su(2)} \rangle
\\\la{hbbsu2}
&&~~~~~~~~~~~~=-
{\tl\ov P_+}\sum_{k\neq j}^M
{n_j n_k +n_k^2 \ov \om_{j}\om_{k}} = -
{\tl\ov 2 P_+}\sum_{k\neq j}^M
{(n_j+n_k)^2 \ov \om_{j}\om_{k}}\,,
\eea
where for simplicity we used the notation $\om_j\equiv\om_{n_j}=\sqrt{1+\tilde\lambda\, n_k^2}$. Now to find the $1/J$ correction 
to the energy of the string state we should solve the equation 
\bea
\label{su2lc}
E - J = \sum_{k=1}^M \sqrt{1 + {4\l n_k^2\ov (E+J)^2}} -
{\tl\ov 2 P_+}\sum_{k\neq j}^M
{(n_j+n_k)^2 \ov \om_{j}\om_{k}}
\eea
in powers of $1/J$ keeping $\l' = \l/J^2$ and $M$ finite. A simple algebra gives
\bea\la{Jcorsu2}
E_{su(2)} - J = \sum_{k=1}^M\bar\om_{k} - {\l'\ov 4J}\sum_{k=1}^M\sum_{j=1}^M
{n_k^2 \bar\om_j^2 + n_j^2\bar\om_k^2\ov \bar\om_k\bar\om_j} -{\l'\ov 4J}\sum_{k\neq j}^M
{(n_j+n_k)^2 \ov \bar\om_{j}\bar\om_{k}}\,,
\eea
where now $\bar\om_k:=\sqrt{1+\lambda'\, n_k^2}$ with the BMN coupling constant
$\lambda':={\lambda}/{J^2}$.
Taking into account the level-matching condition one can easily check that 
(\ref{Jcorsu2}) coincides with the expression obtained in \cite{AFS} by using the quantum string 
Bethe ansatz, and in \cite{Swanson} by using a rather complicated 
string Hamiltonian in the uniform gauge 
$t=\tau$, $p_\p =J$.

Since the quartic Hamiltonian $H_4$ contains terms describing interactions of 
operators $\a_{1,n}^\pm$ with operators charged under other U(1) subgroups, 
the state (\ref{su2state}) gets corrections which depend on these operators. 
We will argue at the end of this section that there is a unitary transformation 
which transforms the Hamiltonian to such a form that the action of the transformed 
Hamiltonian on the states (\ref{su2state}) 
(and in general on states dual to operators from closed sectors) is closed. 
This Hamiltonian is a string 
analog of the field theory dilatation operator, and its restriction to operators 
$\a_{1,n}^\pm$  can be considered as an effective Hamiltonian for the $\su(2)$ sector.

\subsection{$\sl(2)$ sector}
The $\sl(2)$ sector of \N SYM consists of operators of the form
\bea\la{sl2oper}
O_{sl(2)} = \Tr \left( D_-^M Z^J + \mbox{permutations} \right)\,,
\eea
where $D_-$ is the covariant derivative in a light-cone 
direction carrying  unit charge under 
the U(1) subgroup of $SU(2,2)$, that in the string picture corresponds to 
the U(1) generated by $\Phi_1^{AdS}$, 
see appendix A. The operators, correspondingly, carry $J$ and $M$ 
units of the charges.
They are again highest-weight, and string states dual to operators from the $\sl(2)$ sector of \N SYM 
are easily identified 
by analyzing the tables of charges in appendix A. We see that the string state carrying
$M$ units of the charge $S_1$ and having the minimal energy is obtained by 
acting on the vacuum by $M$ creation operators $\b_{1,n}^+$. Therefore, in the BMN limit 
the string states dual to operators from the $\sl(2)$ sector are the states
\bea\la{sl2state}
|\Psi_{sl(2)} \rangle = \b_{1,n_M}^+ \b_{1,n_{M-1}}^+ \cdots \b_{1,n_1}^+  
|0 \rangle\,.
\eea
The computation of the $1/P_+$ correction to this state literally repeats the computation
we did for the $\su(2)$ sector. The only change is the opposite sign of the correction 
(\ref{hbbsu2}), compare (\ref{Hbb}), i.e.~we have
\begin{equation}
\label{sl2lc}
E - J = \sum_{k=1}^M \sqrt{1 + {4\l n_k^2\ov (E+J)^2}} +
{\tl\ov 2 P_+}\sum_{k\neq j}^M
{(n_j+n_k)^2 \ov \om_{j}\om_{k}}\, ,
\end{equation}
and therefore, the $1/J$ correction for the $\sl(2)$ state takes the
form 
\bea 
\la{Jcorsl2} 
E_{sl(2)} - J = \sum_{k=1}^M\bar\om_{k} -
{\l'\ov 4J}\sum_{k=1}^M\sum_{j=1}^M {n_k^2 \bar\om_j^2 +
n_j^2\bar\om_k^2\ov \bar\om_k\bar\om_j} +{\l'\ov 4J}\sum_{k\neq j}^M
{(n_j+n_k)^2 \ov \bar\om_{j}\bar\om_{k}}\,.  
\eea
Again it is straightforward to check that (\ref{Jcorsl2}) coincides
with the expression obtained in \cite{Swanson}.  Let us mention that
the fact that the $1/J$ correction for $\su(2)$ states differs from
the one for $\sl(2)$ states just by a sign of one term seems to not
have been noticed before.  This sign difference between the $\su(2)$ and
$\sl(2)$ reflects the fact that curvatures in the ${\rm S}^5$ and $AdS_5$
parts are equal but of opposite signs.

\subsection{$\su(1|1)$ sector}
The detailed discussion of the $\su(1|1)$ sector in \N SYM and string theory was given 
in \cite{Staudacher:2004tk,Alday:2005jm,AFlc}. We find it useful, for completeness and 
since we have changed the basis of gamma matrices, to 
review shortly the consideration in \cite{Alday:2005jm}. 

The $\su(1|1)$ sector of ${\cal N}=4$ SYM consists of operators of the form
\bea
\label{gssu11}
O_{su(1|1)} = {\rm tr}\big(Z^{J-\frac{M}{2}} \Psi^M  + \mbox{permutations}  
\big)\,.
\eea
The fermion $\Psi$ is the highest
weight component of the gaugino from the vector
multiplet. The gaugino $\Psi_{\a}$ belongs to the vector
multiplet, it is neutral under $\su(3)$ which rotates the three complex
scalars between themselves, and it carries the same charge $1/2$ 
under any of the three $U(1)$  subgroups of SU(4). 
The corresponding
Lie algebra element is
 \bea \Phi_{\underline{\un(1)} \times \su(3) } ={\footnotesize
\left(\begin{array}{rrrr}
i(\xi_1+\xi_2+\xi_3) & 0 ~&~ 0 ~&~ 0 \\
0~&-i\xi_1 ~&~ \alpha_1+i\beta_1 ~&~ \alpha_2+i\beta_2 \\
0~&-\alpha_1+i\beta_1 ~&~ -i\xi_2 ~&~ \alpha_4+i\beta_4  \\
0~&-\alpha_2+i\beta_2 ~&~ -\alpha_4+i\beta_4 ~&~ -i\xi_3 
\end{array}
\right)} \, ,\eea where the $\su(3)$ part is obviously specified by choosing
$\xi_3=-\xi_1-\xi_2$.
It also transforms as a spinor under one
of the $\su(2)$'s from the Lorentz algebra $\su(2,2)$ and is
neutral under the other. Therefore, the highest
weight component $\Psi$ carries the charges $S_1 = 1/2$ and $S_2 = -1/2$.
Therefore, the operators from the $\su(1|1)$ sector have the following charges:
$S_1 = M/2$, $S_2 = -M/2$, $J_1=M/2$, $J_2=M/2$ and $J_3=J$. 

Coming back to string theory we notice that out of 8 fermions $\eta$ and $\theta$ only
$\theta_1$ and $\theta_2$ are neutral under  the $\su(3)$ subgroup, and, therefore, they are dual to 
the components of the gaugino $\Psi_{\a}$. 
From the table of charges in appendix A we see that $\theta_1$ should be identified with the highest
weight component $\Psi$. Thus, in the BMN limit 
the string states dual to operators from the $\su(1|1)$ sector are the states
\bea\la{su11state}
|\Psi_{su(1|1)} \rangle = \theta_{1,n_M}^+ \theta_{1,n_{M-1}}^+ \cdots \theta_{1,n_1}^+  
|0 \rangle\,.
\eea
As was shown in \cite{AFlc}, in the uniform light-cone gauge 
the string theory reduced to the $\su(1|1)$ sector is described 
by a free fermion. We can also see that from our quartic Hamiltonian (\ref{hff}). 
There is, therefore, no $1/P_+$ correction to the free spectrum of the state (\ref{su11state}),
and the $1/J$ correction is just obtained by expanding the frequencies $\om_n$ in powers of 
$1/J$. The result of the simple computation is  \cite{AFlc}
\bea\la{Jcorsu11}
E_{su(1|1)} - J = \sum_{k=1}^M\bar\om_{k} - {\l'\ov 4J}\sum_{k=1}^M\sum_{j=1}^M
{n_k^2 \bar\om_j^2 + n_j^2\bar\om_k^2\ov \bar\om_k\bar\om_j}\,.
\eea
This correction was first computed in \cite{Alday:2005jm} by using the 
uniform gauge $t=\tau$, $p_\p =J$ in which the $\su(1|1)$ sector is described 
by a nontrivial integrable model of an interacting Dirac fermion. 
It was also guessed in \cite{Swanson} by analyzing the known 3-impurity result \cite{Callan}. 

\subsection{$\su(1|2)$ sector}
The $\su(1|2)$ sector can be considered as the union of the $\su(2)$ and $\su(1|1)$ sectors 
\cite{BSsu12},
because it consists of operators of the form
\bea\la{su12oper}
O_{su(1|2)} = \Tr \left( Z^{J-\frac{M}{2}} \Psi^M X^K + \mbox{permutations}
 \right)\,.
\eea
Since we already know that $X$ and $\Psi$ correspond to $\a_1^+$ and $\theta_1^+$, respectively,
string theory states dual to the operators  (\ref{su12oper}) are of the form
\bea\la{su12state}
|\Psi_{su(1|2)} \rangle = |\Psi_{su(1|1)}\rangle\otimes |\Psi_{su(2)} \rangle
= \theta_{1,n_M}^+ \theta_{1,n_{M-1}}^+ \cdots \theta_{1,n_1}^+  \cdot
\a_{1,j_K}^+ \a_{1,j_{K-1}}^+ \cdots \a_{1,j_1}^+  
|0 \rangle\,,~~~~~
\eea
where the mode numbers satisfy the level-matching condition
\bea\nonumber
\sum_{i=1}^M n_{i} + \sum_{m=1}^K j_{m} = 0\,.
\eea
This time, however, there is a mixing of states with the same numbers
$M$ and $K$, because any state obtained from (\ref{su12state}) by a
permutation of the mode numbers $n_i$ and $j_m$ has the same energy as
the state (\ref{su12state}) does.  If all mode numbers are different
the number of all these states is equal to ${(M+K)!\ov M!K!}$, which
makes the problem of computing the $1/P_+$ correction highly
nontrivial. Still the fact that the number of fermions and bosons is
the same for all these states appears to make the problem feasible.

\subsection{$\su(2|3)$ sectors}
The $\su(2|3)$ sector \cite{Beisertsu23} is an extension of the $\su(1|2)$ sector.
It consists of operators of the form
\bea\la{su23oper}
O_{su(2|3)} = \Tr \left( Z^{J-\frac{M_+}{2} -\frac{M_-}{2}}X^{J_1} Y^{J_2}
\Psi_+^{M_+} \Psi_-^{M_-}  + \mbox{permutations} \right)\,,
\eea
where $\Psi_+$ is the highest
weight component of gaugino the $\Psi_{\a}$ from the vector
multiplet that was denoted as $\Psi$ in previous subsections, and 
$\Psi_-$ is the lowest weight component.
String theory states dual to the operators  (\ref{su23oper}) are of the form
\bea\la{su23state}
|\Psi_{su(2|3)} \rangle 
=\theta_{2,n_{M_-}}^+ \cdots \theta_{2,n_1}^+  \cdot 
\theta_{1,m_{M_+}}^+ \cdots \theta_{1,m_1}^+ \cdot
\a_{2,k_{J_2}}^+ \cdots \a_{2,k_1}^+ \cdot
\a_{1,l_{J_1}}^+  \cdots \a_{1,l_1}^+  
|0 \rangle\,.~~~~~
\eea
In the plane-wave limit the space of these string states is highly degenerate. 
Just as it was for string states from the $\su(1|2)$ sector,
we can permute mode numbers of fermions and bosons. Then, a new feature appears 
in the $\su(2|3)$ sector. One can easily check by using the tables of charges 
in appendix A that the operators $\a_{2,n}\a_{1,m}$ and $\theta_{2,k}\theta_{1,l}$ 
have the same charges $S_i$ and $J_i$, and, therefore, we can replace any pair 
of operators $\a_{2,n}$ and $\a_{1,m}$ in the state (\ref{su23state}) by a pair of operators 
$\theta_{2,n}$ and $\theta_{1,m}$ (or $\theta_{2,m}$ and $\theta_{1,n}$) 
with the same mode numbers $n$ and $m$ 
without changing the plane-wave energy of the state. Thus, there is 
a mixing of states with different numbers of bosons and fermions. 
This new feature is a string theory analog of the dynamic nature of the long-range 
spin chain that describes the $\su(2|3)$ sector in \N SYM \cite{Beisertsu23}. 
Let us note, however, that in string theory only the states with the same 
total number of creation operators, $M = J_1+J_2+M_-+M_+$, can mix. 
It seems hardly possible to compute the $1/P_+$ correction to an arbitrary 
$\su(2|3)$ string state by using conventional methods. 

\medskip

String states dual to all the remaining closed sectors of \N SYM can
be also easily identified by using the tables of charges from appendix
A.  Let us also comment that the main obstacle in explicit
computations of energy shift for arbitrary M-excitation state which
originated from the large mixing problem, can be significantly reduced
if one \emph{assumes} quantum integrability of the full model (at
order $1/P_+$). If the system is integrable, it is enough to determine
the energy shift of the arbitrary 3-excitation state. The energy shift
for the M-excitation state is then given by the sum over the corrected
energies of individual impurities (magnons).

\subsection{Effective Hamiltonians for closed sectors}
As was discussed above in this section, the eigenstates of the
complete Hamiltonian depend on all the 8+8 creation operators even for
string states dual to operators from closed gauge theory sectors.  On
the other hand, according to the AdS/CFT correspondence, the string
Hamiltonian should be equivalent to the dilatation operator.  That
means that there should exist  a unitary transformation such that the
transformed Hamiltonian would have properties similar to the ones of
the dilatation operator, in particular, its action on string states
dual to operators from closed sectors would be closed.

We will show here that assuming the finiteness of the quantum theory 
such a unitary transformation exists in perturbation theory around the plane-wave.

In this section we denote the creation and annihilation operators 
as $A^\pm_{a, n}$, where $a$ is an index that distinguishes
 operators of different types,
and $n$ is a mode number. The quantum string Hamiltonian will be of the form
\bea \la{Hc}
H = H_2 + {1\ov P_+}H_4 + {1\ov P_+^2}H_6 + \cdots\,.
\eea
Here the quadratic Hamiltonian $H_2$ reads 
\bea
H_2 = \sum_{a,n}\om_{a,n} A^+_{a, n} A^-_{a, n} \,,
\eea
where the frequencies may in general depend on $a$ and $P_+$
\bea
\om_{a,n} = \om_n + {1\ov P_+}\om_n^{(1)} +  {1\ov P_+^2}\om_n^{(2)} +\cdots\,.
\eea
The quartic Hamiltonian $H_4$ is of the most general form   
\bea\nonumber
H_4 &=& \sum_{a,n;b,m;c,k;d,l} g^{++++}_{a,n;b,m;c,k;d,l}A^+_{a, n}A^+_{b, m}A^+_{c, k}A^+_{d, l} 
+ g^{+++-}_{a,n;b,m;c,k;d,l}A^+_{a, n}A^+_{b, m}A^+_{c, k}A^-_{d, l} + h.c.\\\la{h4c}
&~&~~~~~~~~~~~~~+g^{++--}_{a,n;b,m;c,k;d,l}A^+_{a, n}A^+_{b, m}A^-_{c, k}A^-_{d, l}\,,
\eea
where the coupling constants $g_{a,n;b,m;c,k;d,l}$ may also depend on $1/P_+$. The remaining 
Hamiltonians 
$H_6$ and higher are also assumed to be of the most general form. For simplicity we restrict the consideration
to the quartic Hamiltonian.

First of all we show that there exists a unitary transformation that removes 
all terms
with different numbers of creation and annihilation operators. The construction is perturbative in
$1/P_+$, and the unitary transformation is of the form
\bea\la{untr}
U = e^V\,,\quad V={1\ov P_+}V_4 + {1\ov P_+^2}V_6 +\cdots\,,
\eea
where $V_i$ are polynomials of the $i$-th order in the creation and annihilation operators
\bea\la{V4}\nonumber
V_4 = \sum_{a,n;b,m;c,k;d,l} f^{++++}_{a,n;b,m;c,k;d,l}A^+_{a, n}A^+_{b, m}A^+_{c, k}A^+_{d, l} 
+ f^{+++-}_{a,n;b,m;c,k;d,l}A^+_{a, n}A^+_{b, m}A^+_{c, k}A^-_{d, l} - h.c.\,.
\eea
Under the unitary transformation the Hamiltonian transforms as follows
\bea
H\rightarrow U H U^\dagger\,.
\eea
It is not difficult to see that to remove all unwanted terms from  (\ref{h4c}) 
at the leading order in $1/P_+$ 
we should make the following choice of the constants $f_{a,n;b,m;c,k;d,l}$
\bea\nonumber
f^{++++}_{a,n;b,m;c,k;d,l} = {g^{++++}_{a,n;b,m;c,k;d,l}\ov 
\om_{a,n}+ \om_{b,m}+ \om_{c,k}+ \om_{d,l}}\,,\quad 
f^{+++-}_{a,n;b,m;c,k;d,l} = {g^{+++-}_{a,n;b,m;c,k;d,l}\ov 
\om_{a,n}+ \om_{b,m}+ \om_{c,k}- \om_{d,l}}\,.
\eea
It is important to stress that since in perturbation theory 
$\om_{a,n}+ \om_{b,m}+ \om_{c,k}- \om_{d,l}$ is {\it not} 
equal to 0 for any choice of the mode numbers, the unitary transformation 
is well-defined.\footnote{Strictly speaking, since we are dealing with a system with infinite 
number of degrees of freedom to have a well-defined transformation one should introduce an ultra-violet 
cut-off, e.g. by replacing $\om_{a,n}$ by $\om_{a,n}e^{\epsilon|n|}$ where $\epsilon$ 
is the regularization parameter. In fact, one would need a regularization parameter even to 
normal-order the Hamiltonian.}

Then up to terms of order $1/P_+^2$ the Hamiltonian $H_4$ takes the
form \bea\la{H4tr} H_4 &=& \sum_{a,n;b,m;c,k;d,l}
g^{++--}_{a,n;b,m;c,k;d,l}A^+_{a, n}A^+_{b, m}A^-_{c, k}A^-_{d, l}\,.
\eea The unitary transformation induces additional unwanted terms of
order $1/P_+^2$ but all these terms can be removed by a similar
unitary transformation. The unitary transformations at higher order in
$1/P_+$ will typically also induce corrections to the functions
$g^{++--}_{a,n;b,m;c,k;d,l}$.  It is clear that the procedure can be
carried out to any order in $1/P_+$.  The resulting Hamiltonian hence
contains only terms with an equal number of creation and annihilation
operators, with the coefficients which are functions of $1/P_+$. This
Hamiltonian can be considered as a string analog of the field theory
dilatation operator because as we will see in a moment it maps a state
from a closed sector to another state of this sector.\footnote{Note
that this is not necessarily in contradiction with \cite{Min}, who
argued for the non-perturbative violation of the closedness of the
$\su(2)$ sector, as our argument here is purely pertubative.}

To simplify the notations we concentrate our attention on the $\su(2)$
sector, but the conclusion is valid for all closed subsectors. Let us
assume that we act by the ``diagonal'' Hamiltonian of the form
(\ref{H4tr}) on a state from the $\su(2)$ sector.  Then only the terms
which do not contain any other annihilation operators but $\a_{1,n}^-$
can produce a nontrivial state. All other terms acting on an $\su(2)$
state produce the vacuum.  Finally, taking into account that the
product of $M$ operators $\a_{1,n}^-$ carries the minimal charge $J_1
= -M$, we find that the only way to compensate the charge is to
multiply them by $M$ creation operators $\a_{1,n}^+$.  Thus, the
relevant terms in the Hamiltonian are just obtained by setting all
operators but $\a_{1,n}^\pm$ to zero, and the action of the resulting
Hamiltonian is closed on the $\su(2)$ string states.  This Hamiltonian
can be considered as an effective Hamiltonian for the $\su(2)$ sector,
and its expansion in powers of $\l'$ should reproduce the
Landau-Lifschitz Hamiltonian derived in \cite{Kruc,KT}.

One could try to simplify the Hamiltonian (\ref{H4tr}) by using a unitary 
transformation with $V$ of the form
\bea
V = \sum_{a,n;b,m;c,k;d,l}{g^{++--}_{a,n;b,m;c,k;d,l}\ov 
\om_{a,n}+ \om_{b,m}- \om_{c,k}- \om_{d,l}}A^+_{a, n}A^+_{b, m}A^-_{c, k}A^-_{d, l}\,.
\eea
If the denominator $\om_{a,n}+ \om_{b,m}- \om_{c,k}- \om_{d,l}$ would never vanish then 
we could remove all quartic terms. It is clear, however, that in perturbation theory in $1/P_+$ 
it vanishes if $k=n, l=m$
or  $k=m, l=n$, and therefore these terms cannot be removed. One can show that 
the denominator does not vanish for any other choice of mode numbers, and, therefore, we 
can reduce (\ref{H4tr}) to the following simple form
\bea\la{H4trs}
H_4 &=& \sum_{a,n;b,m;c;d} g^{++--}_{a,n;b,m;c;d}A^+_{a, n}A^+_{b, m}A^-_{c, n}A^-_{d, m}\,.
\eea
A Hamiltonian of this form allows a straightforward computation of the energy 
of string states from rank 1 closed sectors.

\section{Quantum string light-cone Bethe equations}

It has been proposed in
\cite{AFS} that the
energies $E$ of the $\AdS$ quantum string, as measured with respect 
to the global time coordinate $t$, should arise as solutions of a
set of quantum string Bethe equations.
 The energy $E$
is natural from the perspective of comparing with the
dual \emph{gauge} theory and its scaling dimensions.  
However, as we have seen in previous
chapters, the energy in global time $E$ is not the most natural 
quantity for the purposes of the quantization of the $\AdS$
string in light-cone gauge, but rather
the world-sheet Hamiltonian $H_{lc}=E-J$. A natural question
to ask now is, whether it is possible to write down a set of
light-cone quantum string Bethe equations directly yielding the spectrum of
$H_{lc}$. The expectations is, that this set of equations takes a simpler
form than the quantum string Bethe equations of 
\cite{AFS,Staudacher:2004tk,BSsu12} 
directly leading to $E$. 

For this let us now assume that the quantum string is integrable
in the sense that the elementary world-sheet excitations (``magnons'') interact 
with each other only via two body interactions, described by the S-matrix,
as was advocated by Staudacher \cite{Staudacher:2004tk}, whose logic we 
closely follow.
The system should then be described by the fundamental equation 
\begin{equation}
\label{fund}
{P_+ \over 2} p_k = 2 \pi n_k + \sum_{j \neq k}^M \theta(p_k, p_j) \,
,
\end{equation}
where the S-matrix is given by $S(p,q)=\exp[i\,\theta(p,q)]$.
In the plane-wave limit, at leading order in the $1/P_+$ expansion,
the system is \emph{free} (i.e. $\theta(p_j,p_k) = 0$). Thus $p_k$
has the perturbative expansion for $P_+\to\infty$ with $\tilde\lambda$ of 
(\ref{tla}) held fixed
\begin{equation}
\label{expansion}
p_k = {4 \pi \over P_+}  n_k + {\delta p_k\over P_+^2} + 
\mathcal{O}\left( {1\over P_+^3} \right) \,
\end{equation}
where $\delta p_k$ are corrections. From this one determines
 that 
\begin{equation}
\label{shift}
\delta p_k= 2\, P_+\, \sum_{j\neq k}^M\, \theta(\frac{4\pi\, n_k}{P_+},
\frac{4\pi\, n_j}{P_+})\, .
\end{equation}
Next, we take our elementary excitations to satisfy the dispersion relation 
\begin{equation}
\label{dl}
E_{lc}(p_k)=\sqrt{1 + {\lambda \over 4\, \pi^2}\, p_k^2 } \, ,
\end{equation}
and take the total world-sheet energy $H_{lc}$ to be additive 
$E_{lc}=\sum_{k=1}^M E_{lc} (p_k)$. 
Inserting the corrections $\delta p_k$ of (\ref{shift}) and 
(\ref{expansion}) into the dispersion relation (\ref{disprel}), 
we deduce an expression for the shift of $P_-$ at order $1/P_+^2$
in terms of the S-matrix
\begin{equation}
\label{dp}
\delta P_- =  {\tilde\lambda\, P_+  \over 2 \pi} 
\sum_{k,j=1, k \neq j}^M {n_k \over ~ \sqrt{1 + \tilde{\lambda} n_k^2}}~ 
\theta({4 \pi \over P_+}n_k , {4 \pi \over P_+} n_j) \, .
\end{equation}
On the other hand, the semiclassical quantization of the string yielded for the 
$E_{lc}$ shift the equations (\ref{su2lc}) and (\ref{sl2lc}). 
By comparing (\ref{dp}) to (\ref{su2lc}) and
(\ref{sl2lc}), we can extract the S-matrix at leading order $1/P_+^2$
\begin{eqnarray}
\label{lcS}
\theta (p_k, p_j) &=&  -\frac{\sign}{2}\,
{(p_k + p_j)^2 \over p_k \omega_j  - p_j \omega_k} \,
\qquad \mbox{where}\quad 
\begin{cases}
\sign=1 & \mbox{for $\su(2)$} \\
\sign=0 & \mbox{for $\su(1|1)$}\\
\sign=-1 & \mbox{for $\sl(2)$} 
\end{cases}
 \nonumber \\
\omega_k &=& \sqrt{1 + {{\lambda} \over  4\pi^2} p_k^2} \, .
\end{eqnarray}

This near plane-wave S-matrix is singular at $p_k=p_j$. It is clear, however, that this singularity 
is an artifact of the expansion in $1/P_+$ similar to the singularity of the XXX spin chain S-matrix
in the long spin chain length limit. 
It turns out that the fundamental relation (\ref{fund}) with the above
expression for the scattering phase shift $\theta(p_k,p_j)$ follows from
the very compact set of \emph{light-cone} Bethe equations with 
an S-matrix regular at $p_k=p_j$
\begin{equation}
\label{lcBethe}
e^{i\,{P_+\over 2}\,p_k} = \prod_{j=1, \,j \neq k}^M 
\left({x_k^+ - x_j^-  \over x_k^- - x_j^+}\,e^{i(p_j-p_k)}
\right)^\sign  \, ,
\end{equation}
where we have used the common variables $x^\pm_k=x^\pm(p_k)$ with
\begin{equation}
x^{\pm}(p) ={1 \over 4}\, (\cot\frac{p}{2}\pm i)\,
\left(1 + \sqrt{1 + {\lambda \over \pi^2} \sin^2 {p \over 2}} \right) \, ,
\end{equation}
first introduced in \cite{Beisert:2004jw}.
It can be checked that this S-matrix, after rescaling $p_k
\rightarrow 2 p_k/P_+$ and expanding in $1/P_+$, 
indeed reduces to the scattering phase of eq.~(\ref{lcS}).
Moreover, it is then natural to assume a generalized dispersion relation
of the form
\begin{equation}
\label{disprel}
E^{\rm gen}_{lc}(p_k) = \sqrt{1 + {\lambda \over \pi^2}\sin^2 \left( {p_k \over 2}\right)} \, ,
\end{equation}
however the potential sine structure would only manifest itself in the 
next-to-leading corrections to the plane-wave limit of order $1/P_+$.

In addition, we are only considering translationally 
invariant states as a consequence of the level matching condition such
that
\begin{equation}
\label{peri}
\sum_{j=1}^M p_j = 0 \, .
\end{equation}
Using this our light-cone Bethe equations can be rewritten in an even simpler
form as
\begin{equation}
\label{lcBethe2}
e^{i{{P_++\sign M}\over 2}p_k} = \prod_{j=1, \,j \neq k}^M 
\left({x_k^+ - x_j^-  \over x_k^- - x_j^+}\right)^\sign  \, .
\end{equation}
The associated dispersion relation for the world-sheet energy $E_{lc}$ is stated
in (\ref{disprel}).

The light-cone Bethe equations can be used to compute sub-leading
$1/P_+^2$ corrections.  We do not expect, however, that they will
produce the correct result.  In particular, the anomaly computation of
\cite{ZamaSchaefer}, observed a discrepancy between the quantum string
Bethe predictions and the results of the semiclassical
string quantisation. The anomaly in \cite{ZamaSchaefer} arose purely
from the short distance behavior of the term on the RHS of
(\ref{lcBethe2}), while the other terms which are different in (\ref{lcBethe2}) with respect to equations of (\cite{AFS}), did not contribute.  Thus, the anomaly
prediction from (\ref{lcBethe2}) will be the same as those in
\cite{ZamaSchaefer}, and hence will not cure the problem which
(\cite{AFS}) faced. 




The Bethe equations (\ref{lcBethe}) have been derived by using the expansion around the plane wave.
On the other hand 
one should expect them to also reproduce the leading $\l^{1/4}$ asymptotic behavior 
of short strings in the strong coupling (flat-space) limit and 
the finite-gap integral equations of \cite{Kazakov:2004qf, Kazakov:2004nh} 
which describe
strings spinning in $R\times S^3$ and ${\rm AdS}_3 \times R$ in the scaling 
 limit of \cite{FTspin}, just as the quantum string Bethe equations of \cite{AFS} did.

\subsection{Strong coupling limit}

The leading $\l^{1/4}$ asymptotic behavior of short strings in the strong coupling (flat-space) limit
was discussed in detail for the $\su(1|1)$ sector in \cite{AFlc}. The consideration there
is also valid for the $\su(2)$ and $\sl(2)$ sectors because as one can easily see the S-matrix
does not contribute in the strong coupling limit $\l\to\infty\,,\ \l/P_+^4$ fixed:
The quasi-momenta then have an expansion of the form
\begin{equation}
p^{\rm s. c.}_k= \frac{p^0_k}{\lambda^{1/4}} + \frac{p^1_k}{\lambda^{1/2}}
+\ldots\, .
\end{equation}
Then obviously the scattering phase of (\ref{lcS}) scales as 
$\theta(p_k,p_j)\sim \frac{1}{\sqrt{\lambda}}$ in consequence of $\omega(p_k)\sim
\lambda^{1/4}$. Therefore the fundamental equation (\ref{fund}) yields
\begin{equation}
p^0_k=\frac{4\pi\,  n_k}{E^{(0)}} \qquad \mbox{where} \qquad
P_+= E+J= \lambda^{1/4} \, E^{(0)} + E^{(1)} + J + {\cal O}(\lambda^{-1/4})\, .
\end{equation}
Plugging this into the dispersion relation (\ref{dl}) we find
\begin{equation}
\label{diese}
E_{lc}=\lambda^{1/4} \, E^{(0)} + {\cal O}(1) = \sum_{k=1}^M\, \lambda^{1/4}\, 
\frac{2 \, |n_k|}{E^{(0)}} + {\cal O}(1)\, .
\end{equation}
Due to the level matching condition $\sum_kn_k=0$ and one may define
the level number $n$ as the sum over the positive $n_k$. This implies
$\sum_k|n_k|=2\, n$. With this definition
we indeed reproduce the result of \cite{Gubser:1998bc} upon solving the
quadratic equation (\ref{diese}) for $E^{(0)}$
\begin{equation}
E=2\,\lambda^{1/4}\,\sqrt{n} + {\cal O}(1)\,.
\end{equation}
Hence in the light-cone gauge the leading $\l^{1/4}$ asymptotics just comes from 
the spectrum of the free harmonic oscillator.

\subsection{Spinning string limit}

\newcommand{\contourgauge}{\mathbf{C}}
\newcommand{\pint}{\makebox[0pt][l]{\hspace{2.6pt}$-$}\int}

To derive the integral equations in the spinning string limit
it is convenient to use the logarithmic form (\ref{fund}) of
the Bethe equations (\ref{lcBethe}). To take the limit we rescale
momenta as $p_k \rightarrow 2 p_k /P_+ $ and introduce the
distributional density \cite{BMSZ}
\bea
\label{den}
\rho(p)=\frac{2}{P_+}\sum_{k=1}^M\delta(p-p_k)\, ,\qquad
\int_{\contourgauge} dp\ \rho (p ) =\frac{2M}{P_+}\, .
\eea
Then in the limit $P_+\to\infty\,,\ M/P_+$ fixed we get the following integral equation
\bea\la{e1}
p = \pint_{\contourgauge} dq \rho(q) \theta (p, q) \,,
\eea
where
\bea\la{e2}
\theta (p, q) = -\frac{\sign}{2}\,
{\left(p + q\right)\left(p\, \omega(q)  + q\, \omega(p)\right) \over p  - q } \, ,\qquad
\omega(p) = \sqrt{1 + {\tilde{\lambda} \over  4\pi^2} p^2} \, .
\eea
This equation should be supplemented by the zero-momentum condition
\bea\la{e3}
\int_{\contourgauge} dp\ \rho (p )\, p = 0\,.
\eea
Then, solving eqs.(\ref{den}), (\ref{e1}) and (\ref{e3}), we can find
the light-cone energy of a spinning string by using the equation
\bea\la{e4}
\frac{2\, E_{lc}}{P_+} = \int_{\contourgauge} dp\ \rho (p )\,\om(p)\,.
\eea

\medskip

To compare equations  (\ref{den}), (\ref{e1}), (\ref{e3}) and (\ref{e4})
with the finite-gap integral equations
of \cite{Kazakov:2004qf, Kazakov:2004nh} we should start with their unscaled
form\footnote{We restrict our attention
to the $\su(2)$ sector because
the consideration of the equations in
the $\sl(2)$ sector literally repeats the one we do for the $\su(2)$ sector.
}
\bea
\la{kmmz}
2 \pint_{\contourgauge} \,dy\, {\rho_s(y) \over x - y} &=&
{2 \pi (P_++P_-) \over \sqrt{\lambda}} {x\over x^2 -{\tl \ov 16\pi^2}}\,,\\
\la{bc2}
\int_{\contourgauge} \,dx\, {\rho_s(x) \over x} &=& 2 \pi m =0\,, \\
\label{bc1}
\int_{\contourgauge} \,dx\, \rho_s (x) &=& {2 \pi \over \sqrt{\lambda}} (P_- +  M ) \,, \\
\label{bc3}
\int_{\contourgauge} \,dx\, {\rho_s(x) \over x^2} &=& {2 \pi \over \sqrt{\lambda}} (P_- - M) \,  ,
\end{eqnarray}
where we set the winding number to 0 because our Bethe equations have been
derived under this assumption.

First, we rescale the spectral parameter as $x\rightarrow 2 \pi P_+ x /
 \sqrt{\lambda} $. This rescaling should be contrasted to the rescaling
$x\rightarrow 4 \pi (J+M) x / \sqrt{\lambda} $
performed in \cite{Kazakov:2004qf}, in order to achieve the
comparison with the gauge theory Bethe ansatz \cite{BSsu12}.

Then, it is easy to see that after the rescaling, the set of equations (\ref{kmmz}), (\ref{bc2}),
(\ref{bc1}) and (\ref{bc3}) can be written in the form
\bea
\la{bc2s}
&&\int_{\contourgauge} \,dx\, {\rho_s(x) \over x} = 0\,, \\
\la{bc3s}
&&\int_{\contourgauge}\, dx \,\rho_s(x) \left( 1 - {\tilde{\lambda} \over 16 \pi^2} {1\over x^2}\right)
= {2 M\over P_+}\,,\\
\la{bc1s}
&&{2 P_-\over P_+}=\int_{\contourgauge}\, dx\, \rho_s(x) \left( 1 + {\tilde{\lambda} \over 16 \pi^2}
{1\over x^2}\right)\,,\\
\la{kmmzs}
&&{x\over x^2 -1} =  2\pint_{\contourgauge}\, dy\, {\rho_s(y) \over x - y}\, -\,
{P_-\ov P_+}\,{x\over x^2 -1}\,.\\
\eea
Finally, making the change of the spectral parameter $x$
\bea
\label{TD1}
p = {x \over x^2 - {\tilde{\lambda} \over 16 \pi^2}} \, , \qquad
\rho(p) = {\rho_s(x)\ov p^2\om(p)}\,,
\eea
we find that the finite-gap integral equations (\ref{kmmzs}), (\ref{bc2s}),
(\ref{bc1s}) and (\ref{bc3s}) coincide with the equations
(\ref{den}), (\ref{e1}), (\ref{e3}) and (\ref{e4}) we derived in the spinning string limit.

\section{Conclusions and Outlook}

The bulk of this paper consisted in a rather laborious derivation of
the exact uniform light-cone and kappa-symmetry gauge fixed Lagrangian
in a first order formalism. This enables one  to read off the
(involved) Poisson structure and form of the light-cone Hamiltonian
$H_{lc}=-p_-$ of the $\AdS$ superstring.  We then went on to quantize
this system in the near plane wave limit of taking the constant
light-cone momentum $P_+$ to infinity while keeping
$\tilde\lambda=4\lambda/P_+^2$ fixed. In this limit we could
systematically expand the Hamiltonian to quartic order in physical
fields and study the leading energy shifts of the closed rank-1
subsectors for general states, reproducing the results of Callan
et.~al.~\cite{Callan} in a rather economical fashion. Furthermore we
proved the existence of effective Hamiltonians in closed
subsectors of the theory, which are the analogues of the dilatation operators in the
closed subsectors of the gauge theory. Finally we were able to write
down a novel, compact set of light-cone Bethe equations, which captures
the leading $1/P_+$ energy shifts in the rank-1 subsectors of the
quantum superstring. This set of quantum string equations was shown to
possess the correct strong coupling and spinning string limits known
in the literature.

There are numerous extensions of the present work. First of all the
next order energy shift computation for $H_{lc}$ in the $1/P_+$
expansion is now within reach and should be performed. It will be able
to test the sine structure in the dispersion relation for the
energy. Here subtle quantum ordering ambiguities have to be overcome:
at the leading $1/P_+$ order our simple normal ordering prescription
was justified by the requirement of having an unmodified quadratic
Hamiltonian. At the next-order such a simple normal-ordering
prescription fails, as it would lead to energy shifts of protected
states. We believe that imposing the closure of the algebra of the
system will pave us the correct path to resolve these ordering
ambiguities at the quantum level. We are presently investigating this issue.

The uniform light-cone gauge Hamiltonian we have established should be
the basis of an investigation of the near flat-space limit of $\AdS$.
Here the subtlety lies in the treatment of the zero-modes, which in a
naive approach leads to a breakdown of the perturbation theory about
the flat space point.

Another obvious question is how to generalize our findings to the case
of non-vanishing winding numbers. This should be relevant for the
study of $\gamma$-deformed models \cite{LM,F} which are known to be
related to Green-schwarz superstrings in $\AdS$ subject to twisted
boundary conditions \cite{F,AAFg}.  The twisting effectively
corresponds to a non-vanishing (and non-integer) mode number in the
expansion around plane-wave limits. In the $\gamma$-deformed case
there are (at least) two inequivalent plane-wave limits
\cite{LM},\cite{F}, \cite{gamma} and the light cone type gauges seem
to be very convenient to study $1/J$ corrections to energies of string
states in these limits.

One might also try to find explicit string solutions of our exact
gauge fixed Lagrangian and study quantum fluctuations about these.  In
particular finding circular string would be interesting. It would
allow for a computation of the $1/J^2$ corrections to the energies of
such circular strings that could be potentially simpler than the
computations in the static gauge \cite{FTspin} used before.

Our newly proposed quantum string light-cone Bethe equations call for
a number of checks and extensions.  Their generalization to
the full $PSU(2,2|4)$ structure seems obvious to guess in view of
\cite{BSsu12}. This could be checked by a parallel independent
computation of the energy shifts in larger subsectors based on our
quartic Hamiltonian. 
Finally, we have seen that
focusing on the computation of the corrections to light-cone energy
$\Delta-J$ instead of the dilatation operator $\Delta$ has lead to the
simplified Bethe equations on the string side: the dressing factor has
``disappeared''. It would be very interesting to understand what are
the corresponding changes for the (asymptotic) Bethe equations in the
dual gauge theory, once they are rephrased in the language of the
light cone variables.

We intend to return to some of these questions in future works.

\section*{Acknowledgements}

We wish  to thank Gleb Arutyunov, Niklas Beisert, Matthias Staudacher and  
Arkady Tseytlin
for very useful discussions and comments.
The work of J.~P. is funded by the Volkswagen Foundation.
He also thanks the Max-Planck-Institute for Gravitational Physics
for hospitality.
The
work of S.~F. and M.~Z. was supported in part by 
the EU-RTN network {\it Constituents, Fundamental
Forces and Symmetries of the Universe} (MRTN-CT-2004-005104).

\appendix
\section{Dirac matrices}

Throughout the paper we will use the following explicit representation
of Dirac matrices, 
\begin{eqnarray}
\nonumber \gamma^1&=&{\footnotesize\left(
\begin{array}{cccc}
  0 & 0 & 0 & -1 \\
  0 & 0 & 1 & 0 \\
   0 & 1 & 0 & 0 \\
   -1 & 0 & 0 & 0
\end{array} \right)},\hspace{0.3in}
\gamma^2={\footnotesize\left(
\begin{array}{cccc}
  0 & 0 & 0 & i \\
  0 & 0 & i & 0 \\
   0 & -i & 0 & 0 \\
   -i & 0 & 0 & 0
\end{array} \right)},\hspace{0.3in}
\gamma^3={\footnotesize\left(
\begin{array}{cccc}
  0 & 0 & 1 & 0 \\
  0 & 0 & 0 & 1 \\
  1 & 0 & 0 & 0 \\
  0 & 1 & 0 & 0
\end{array} \right)}, \\
\nonumber 
\gamma^4&=&{\footnotesize \left(
\begin{array}{cccc}
  0 & 0 & -i & 0 \\
  0 & 0 & 0 & i \\
   i & 0 & 0 & 0 \\
   0 & -i & 0 & 0
\end{array} \right)},\hspace{0.28in}~
\gamma^5={\footnotesize \left(
\begin{array}{cccc}
  1 & 0 & 0 & 0 \\
  0 & 1 & 0 & 0 \\
   0 & 0 & -1 & 0 \\
   0 & 0 & 0 & -1
\end{array} \right)} = \Sigma\, .
\end{eqnarray}
satisfying the SO(5) Clifford algebra
$$
\gamma^a\gamma^b+\gamma^b\gamma^a=2\delta_{ab}\, .
$$
Moreover, all of them are hermitian, so that $i\gamma_{a}$ belongs
to $\su(4)$. 

It is also useful to introduce such a basis that all 
bosonic fields have definite 
charges under the U(1) subgroups of PSU(2,2$|$4). We also describe a 
convenient parametrization of 
the fermionic matrices $\chi$ and $\T$.

The complex fields carrying definite charges are 
\bea\la{ll1}
Z_1 &=& z_2 + iz_1\, ;\quad Z_2 = z_4 + iz_3\, ;\quad 
Z_{\bar 1} = z_2 - iz_1\, ;\quad Z_{\bar 2} = z_4 - iz_3\, ;
\\\nonumber
Y_1 &=& y_2 + iy_1\, ;\quad Y_2 = y_4 + iy_3\, ;\quad 
Y_{\bar 1} = y_2 - iy_1\, ;\quad Y_{\bar 2} = y_4 - iy_3\,.
\eea
We want to have the identity
\bea\la{ll2}
z_i\g_i = Z_a \Gamma_a = Z_1 \Gamma_1 + Z_{\bar 1} \Gamma_{\bar 1}+
Z_2 \Gamma_2 + Z_{\bar 2} \Gamma_{\bar 2}\; ,
\eea
which lets us introduce 
\bea\la{ll3}
\Gamma_1 = {1\ov 2}(\g_2 -i\g_1) = {\footnotesize\left(
\begin{array}{cccc}
  0 & 0 & 0 & i \\
  0 & 0 & 0 & 0 \\
   0 & -i & 0 & 0 \\
   0 & 0 & 0 & 0
\end{array} \right)}\, , \quad
\Gamma_{\bar 1}  = \G_1^\dagger ={1\ov 2}(\g_2 +i\g_1) =  {\footnotesize\left(
\begin{array}{cccc}
  0 & 0 & 0 & 0 \\
  0 & 0 & i & 0 \\
   0 & 0 & 0 & 0 \\
   -i & 0 & 0 & 0
\end{array} \right)}\nonumber\\ 
\Gamma_2 = {1\ov 2}(\g_4 -i\g_3) = {\footnotesize\left(
\begin{array}{cccc}
  0 & 0 & -i & 0 \\
  0 & 0 & 0 & 0 \\
   0 & 0 & 0 & 0 \\
   0 & -i & 0 & 0
\end{array} \right)}\, , \quad
\Gamma_{\bar 2}  = \G_2^\dagger ={1\ov 2}(\g_4 +i\g_3) =  {\footnotesize\left(
\begin{array}{cccc}
  0 & 0 & 0 & 0 \\
  0 & 0 & 0 & i \\
   i & 0 & 0 & 0 \\
   0 & 0 & 0 & 0
\end{array} \right)} \, .
\eea
In what follows we will often use for the indices $\bar{1}$ and $\bar{2}$ the following convention:
\bea
\bar{1}\equiv 4\, ,\quad  \bar{2}\equiv 3\, ,\quad\Rightarrow\quad \bar{a}\equiv 5-a\, ,
\quad a=1,2,3,4.
\eea
In particular that means that 
$$
\G_{\bar 1} \equiv \G_4\, ,\quad \G_{\bar 2} \equiv \G_3\, ,\quad \G_{\bar a} \equiv \G_{5-a} 
\, ,\quad \G_{a}^\dagger \equiv \G_{5-a} \,.  
$$
It is also useful to introduce the following two orthogonal projectors
\bea
\la{ll5}
\PP_+ = {1\ov 2}(I_4 +\S) = \left(
\begin{array}{cc}
  I_2 & 0 \\
  0 & 0 
\end{array} \right)\, , \quad \PP_- = {1\ov 2}(I_4 -\S) = \left(
\begin{array}{cc}
  0 & 0 \\
  0 & I_2
\end{array} \right)\, .
\eea
We can write the fermionic $8\times 8$ matrix $\chi$ in the form 
\bea
\chi = \left(
\begin{array}{cc}
  0 & \Theta \\
  \Theta_* & 0 
\end{array} \right) = \s_+\otimes \Theta + \s_-\otimes \Theta_*
\eea
where 
\bea
\s_+ =  \left(
\begin{array}{cc}
  0 & 1 \\
  0 & 0 
\end{array} \right)\, ;\quad \s_- =  \left(
\begin{array}{cc}
  0 & 0 \\
  1 & 0 
\end{array} \right)\, ,
\eea
are the two nilpotent 2 by 2 matrices. 
Note that the 2-dimensional projectors $P^\pm_2$ are expressed 
through $\s^\pm$ as follows
\bea
P^+_2=\s_+ \s_- = \left(
\begin{array}{cc}
  1 & 0 \\
  0 & 0 
\end{array} \right)\, ;\quad P^-_2=\s_- \s_+ = \left(
\begin{array}{cc}
  0 & 0 \\
  0 & 1 
\end{array} \right)\, .
\eea
Then, the kappa-fixed $\T$ can be expanded as follows
\bea\la{ll6}
\Theta &=& \PP_+\eta + \PP_-\td\, ,\quad \eta= \eta_a\G_a \, ,\quad \theta= \theta_a\G_a  \\\nonumber
\Td &=& \eta^\dagger P_+ + \theta P_- 
= P_- \eta^\dagger + P_+\theta\\\nonumber
\Theta_* &=& -\Td \S = -\eta^\dagger \PP_+ + \theta \PP_- 
= -\PP_- \eta^\dagger + \PP_+\theta
\eea
The fermions $\theta_{ij}$ are related to $\eta_a$ and $\theta_a$ as follows
\bea\la{ll8}
&&\theta_{13}=-i\eta_2 \, ;\quad \theta_{14}=i\eta_1 \, ;\quad \theta_{23} =i\eta_4\, ;\quad 
\theta_{24} =i\eta_3\, ; \\\nonumber
&&\theta_{31}=i\td_2  \, ;\quad \theta_{32} =-i\td_4\, ;\quad \theta_{41}=-i\td_1\, ;\quad 
\theta_{42} =-i\td_3\, .
\eea
Since $\S$ anticommutes with all $\G_a$ it also anticommutes with the kappa-gauge fixed 
$\T$ and $\Ts$. 
By using this property one can easily show that
\bea\nonumber
\S_+\chi = -\chi\S_+\,,\quad \S_-\chi = \chi\S_-\,.
\eea

\medskip

The bosonic fields and fermions are charged under the four U(1) subgroups 
of ${\rm SU}(2,2)\times {\rm SU}(4)$ generated by 
\bea\la{uones}
\Phi^{AdS}_i = P^+_2\otimes \Phi_i\,,\quad \Phi^{S}_i = P^-_2\otimes \Phi_i\,,
\eea
where 
$$
\Phi_1 = {\footnotesize
\frac{1}{2}\left(\begin{array}{rrrr}
1 ~&~ 0 ~&~  0 ~&~  0 \\
0 ~&~ -1 ~&~ 0 ~&~  0 \\
 0~&~  0 ~&~ 1 ~&~  0 \\
 0 ~&~  0 ~&~ 0 & -1
\end{array}
\right) }
\, ,\qquad \Phi_2 = {\footnotesize
\frac{1}{2}\left(\begin{array}{rrrr}
1 ~&~ 0 ~&~  0 ~&~  0 \\
0 ~&~ -1 ~&~ 0 ~&~  0 \\
 0~&~  0 ~&~ -1 ~&~  0 \\
 0 ~&~  0 ~&~ 0 & 1
\end{array}
\right) }
\, .
$$
It is not difficult to check (see \cite{Alday:2005jm} for details)  that 
$\Phi^{AdS}_1$ and $\Phi^{AdS}_2$ generate rotations in the planes $z_2z_1$  and $z_4z_3$, 
or multiplication by a phase of $Z_1$ and $Z_2$, respectively. 
We denote these ${\rm AdS}_5$ charges as $S_i$. 
Similarly, 
$\Phi^{S}_1$ and $\Phi^{S}_2$ generate rotations in the planes $y_2y_1$  and $y_4y_3$, 
or multiplication by a phase of $Y_1$ and $Y_2$, respectively. 
We denote these ${\rm S}^5$ charges as $J_i$.

In the tables below we list the fields charges under the four U(1) subgroups. 
Let us stress again that all the fields are neutral under the two U(1) subgroups that 
correspond to shifts of the global time coordinate, and the ${\rm S}^5$ angle $\phi$.

Charges of bosonic fields and creation and annihilation operators:
$$
\begin{array}{|c|c|c|c|c|}
\hline
&S_1&S_2&J_1&J_2\\
\hline
~Z_1,\, P^z_1,\, \b_{1,n}^+,\, \b_{4,n}^-~&+1&~0&~0&~0\\
\hline
~Z_2,\, P^z_2,\, \b_{2,n}^+,\, \b_{3,n}^-~&~0&+1&~0&~0\\
\hline
~Z_3,\, P^z_3,\, \b_{3,n}^+,\, \b_{2,n}^-~&~0&-1&~0&~0\\
\hline
~Z_4,\, P^z_4,\, \b_{4,n}^+,\, \b_{1,n}^-~&-1&~0&~0&~0\\
\hline
\end{array}\,,\quad \begin{array}{|c|c|c|c|c|}
\hline
&S_1&S_2&J_1&J_2\\
\hline
~Y_1,\, P^y_1,\, \a_{1,n}^+,\, \a_{4,n}^-~&~0&~0&+1&~0\\
\hline
~Y_2,\, P^y_2,\, \a_{2,n}^+, \,\a_{3,n}^-~&~0&~0&~0&+1\\
\hline
~Y_3,\, P^y_3,\, \a_{3,n}^+, \,\a_{2,n}^-~&~0&~0&~0&-1\\
\hline
~Y_4,\, P^y_4,\, \a_{4,n}^+,\, \a_{1,n}^-~&~0&~0&-1&~0\\
\hline
\end{array}\,,
$$
Charges of fermions and creation and annihilation operators:
$$
\begin{array}{|c|c|c|c|c|}
\hline
&S_1&S_2&J_1&J_2\\
\hline
~\theta_1,\,\td_4,\, \theta_{1,n}^+,\, \theta_{4,n}^-~&~+{1\ov 2}&~-{1\ov 2}&~+{1\ov 2}&~+{1\ov 2}\\
\hline
~\theta_2,\,\td_3,\, \theta_{2,n}^+,\, \theta_{3,n}^-~&~-{1\ov 2}&~+{1\ov 2}&~+{1\ov 2}&~+{1\ov 2}\\
\hline
~\theta_3,\,\td_2,\, \theta_{3,n}^+,\, \theta_{2,n}^-~&~+{1\ov 2}&~-{1\ov 2}&~-{1\ov 2}&~-{1\ov 2}\\
\hline
~\theta_4,\,\td_1,\, \theta_{4,n}^+,\, \theta_{1,n}^-~&~-{1\ov 2}&~+{1\ov 2}&~-{1\ov 2}&~-{1\ov 2}\\
\hline
\end{array}\,,\quad 
\begin{array}{|c|c|c|c|c|}
\hline
&S_1&S_2&J_1&J_2\\
\hline
~\eta_1,\,\etad_4,\, \eta_{1,n}^+,\, \eta_{4,n}^-~&~+{1\ov 2}&~+{1\ov 2}&~+{1\ov 2}&~-{1\ov 2}\\
\hline
~\eta_2,\,\etad_3,\, \eta_{2,n}^+,\, \eta_{3,n}^-~&~+{1\ov 2}&~+{1\ov 2}&~-{1\ov 2}&~+{1\ov 2}\\
\hline
~\eta_3,\,\etad_2,\, \eta_{3,n}^+,\, \eta_{2,n}^-~&~-{1\ov 2}&~-{1\ov 2}&~+{1\ov 2}&~-{1\ov 2}\\
\hline
~\eta_4,\,\etad_1,\, \eta_{4,n}^+,\, \eta_{1,n}^-~&~-{1\ov 2}&~-{1\ov 2}&~-{1\ov 2}&~+{1\ov 2}\\
\hline
\end{array}\,,
$$
\medskip

Below we collect some useful identities:
\begin{eqnarray}
\label{K1}
&&K_{4,8}^2 = - I_{4,8}  \quad \Sigma_\pm^2 = I  \nonumber\\
&&K_4 \Sigma K_4 = - \Sigma \, , \quad K_8 \Sigma_\pm  K_8 = - \Sigma_\pm 
\nonumber\\
&&K_4 \gamma^i K_4  = - (\gamma^i)^t  \nonumber\\
&& K g^t K = - g \, ,  \quad \quad K (g^{-1})^t K = - g^{-1} \, , \quad \quad K 
\partial_\alpha g^t K = - \partial_\alpha g  \nonumber\\
&& g \Sigma_\pm g^{-1} = g^2 \Sigma_\pm = \Sigma_\pm g^{-2} \, .
\end{eqnarray}

\section{Computing the gauge-fixed Lagrangian}
In this appendix we simplify various terms appearing in the Lagrangian (\ref{Lang2}),
solve the Virasoro constraints and find $x_-'$ and $\bp_-$. 

\subsection{Simplifying ${\bf p_-}$}
Taking into account the decomposition (\ref{bpexp}) of $\bp$, we rewrite ${\bf p_-}$ (\ref{bpm})
in the form
\begin{eqnarray}\la{bpm2}
{\bf p_-}& =& {\bp_- \over 8} \text{Str}\left(\Sigma_8\, (1+2\chi^2)\, g(x)^2 \right)  
-{\bp_+ \over 8} \text{Str}\left((1+2\chi^2)\, g(x)^2 \right) \\\nonumber &+&
{i\bp_M \over 4} \text{Str}\left(\Sigma_M \Sigma_+\, g(x)\, (1+2\chi^2)
\, g(x) \right) 
\end{eqnarray}
where $\Sigma_8 = - \Sigma_+\Sigma_-=\left(
\begin{array}{cc}
I_4 & 0 \\
0 & -I_4
\end{array}\right)$.

In what follows we find it convenient to use the following definitions
\bea\la{g}
g(x) &=& g_+ I_8 + g_-\Sigma_8 + g_M \Sigma_M\, ,\quad 
g_M = \left\{g_a\, ,g_s\right\} \, ,\\\nonumber
g_\pm &=&{1\ov 2}\left( {1\ov \sqrt{1-{z^2\ov 4}}} \pm  {1\ov \sqrt{1+{y^2\ov 4}}}\right)\, ,\quad 
g_a = {z_a\ov 2\sqrt{1-{z^2\ov 4}}}\, ,\quad 
g_s = {y_s\ov 2\sqrt{1+{y^2\ov 4}}} \, ;
\eea
\bea\la{gg}
g(x)^2 &=& G_+I_8 +G_-\S_8 +G_M\S_M\, ,\quad G_M = \left\{G_a\, ,G_s\right\} \, ,\\\nonumber
G_\pm &=&{1\ov 2}\left( {1+{z^2\ov 4}\ov 1-{z^2\ov 4}} \pm  {1-{y^2\ov 4}\ov 1+{y^2\ov 4}}\right)
\, ,\quad 
G_a = {z_a\ov 1-{z^2\ov 4}}\, ,\quad 
G_s = {y_s\ov 1+{y^2\ov 4}} \, .
\eea
Then by using (\ref{p+explicit}) and the properties of Dirac matrices, one can easily get 
\bea\la{pm1B}
{\bf p_- }
&=&-{G_-\ov G_+}P_+ + {G_+^2-G_-^2\ov G_+}\bp_--
\frac{P_+}{4}\str\left( \chi^2\right)\\\nonumber
&~&~~~~~~~~+
 \frac{i}{2}g_N\bp_M\str\left(\left[\S_N,\Sigma_M\right]\,\chi^2\left(
\S_+g_+  -\S_-g_- \right) \right) \,.
\eea
As one can see, ${\bf p_- }$ has only an explicit quadratic dependence on the fermion $\chi$. 
However, the true
 $\chi$ dependence is much more complicated because $\bp_-$ nontrivially depends on 
$\chi$.

\subsection{Simplifying $-\str\,\bp A_{even}^\perp$}
Let us introduce the even and odd components of $g^{-1}(\chi)\pa_\a g(\chi)$
\bea\la{BF}
g^{-1}(\chi)\pa_\a g(\chi) &=& B_\a + F_\a\,,\\\nonumber
B_\a &=&  -{1\ov 2}\chi\pa_\a\chi + {1\ov 2}\pa_\a\chi\chi + 
{1\ov 2}\sqrt{1+\chi^2}\pa_\a\sqrt{1+\chi^2} -
{1\ov 2}\pa_\a\sqrt{1+\chi^2}\sqrt{1+\chi^2}\, ,\\\nonumber
F_\a&=& 
  \sqrt{1+\chi^2}\pa_\a\chi - \chi\pa_\a\sqrt{1+\chi^2}\, .
\eea
Then $A_{even}^\perp$ can be written in the form 
\bea
A_{even}^\perp = -g^{-1}(x)B_\tau g(x)-g^{-1}(x)\pa_\tau g(x)\,.
\eea
The supertrace of the second term with $\bp$ can be easily computed, and we get
\bea
\str\,\bp g^{-1}(x)\pa_\tau g(x) = p_M\pa_\tau x_M\,,
\eea
where the momenta conjugate to the coordinates $x_M =\{z_a, y_s\}$ are given by
\bea
p_M =\{p^z_a, p^y_s\}\,,\quad 
p_a = {\bp_a\ov 1 - {z^2\ov 4}}\, ;\quad p_s = {\bp_s\ov 1 + {y^2\ov 4}}\,.
\eea

\medskip

The term dependent on $B_\a$ can be also written in a more explicit form:
\bea\nonumber
\str\left(\bp g^{-1} B_\a g\right) 
=\frac{i\bp_+}{4}\str\left(\Sigma_+B_\a g^2\right)+
\frac{i\bp_-}{4}\str\left(\Sigma_-B_\a g^2\right)
+\frac{\bp_M}{2}\str\left(\Sigma_M g^{-1}B_\a g\right)
\eea 
Now, taking into account the formulas
\bea\la{l9}
\str\left(\Sigma_+B_\a\right) = 
- \str\left(\Sigma_+\chi\pa_\a\chi\right) \,,\quad
\str\left(\Sigma_-B_\a\right) = 0\,,\quad\str\left(\Sigma_+\chi^n\right) =0\, ,
\eea
that follow from the fact that $\Sigma_+$ anticommutes with $\chi$,  
$\Sigma_-$ commutes with $\chi$ and the explicit form of $B_\a$ we get 
\bea\la{ba}
\str\left(\bp g^{-1} B_\a g\right)=-\frac{i}{4}P_+\,\str\left(\Sigma_+\chi\pa_\a\chi\right)
+\frac{1}{2}\bp_M\,\str\left(g\,\Sigma_M g^{-1}B_\a\right)
\, .
\eea 
The last term in (\ref{ba}) can be also simplified by using the explicit formula (\ref{g}) 
for $g(x)$, commutativity of $\S_\pm$ with $B_\a$, and the properties of gamma matrices. 

The final expression for $-\str\,\bp A_{even}^\perp$ then takes the form
\bea\nonumber
-\text{Str}\left(\bp\,A_{even}^\perp\right) 
&=& p_M\pa_\tau x_M - 
\frac{i}{4}P_+\,\str\left(\Sigma_+\chi\pa_\tau\chi\right)\\\la{l12}
&+&\frac{1}{2}g_N\bp_M\,\str\left(\left[\S_N,\Sigma_M\right] B_\tau\right)\, .
\eea

\subsection{Solving the Virasoro constraint $C_2$: Str$\,\bp\, A^{(2)}_1 = 0$}
The constraint $C_2 = 0$ can be now easily solved to find $x_-' \equiv \pa_1 x_-$ 
\bea
\nonumber
-\text{Str}\left(\bp\,A^{(2)}_1\right) 
= P_+x_-' + p_Mx_M' + 
\frac{i}{4}P_+\,\str\left(\Sigma_+B_1\right)
+\frac{1}{2}\bp_M\,\str\left(g\,\Sigma_M g^{-1}B_1\right) = 0\, .
\eea
Thus
\bea
\la{l13}
x_-' = - {1\ov P_+}\left( p_Mx_M' -
\frac{i}{4}P_+\,\str\left(\Sigma_+\chi\chi'\right)
+\frac{1}{2}g_N\bp_M\,\str\left(\left[\S_N,\Sigma_M\right] B_\s\right) \right)\, .
\eea
The nice feature is that in the light-cone 
gauge $x_-'$ has no dependence on $\bp_-$. 

\subsection{Level-matching condition}

Integrating (\ref{l13}) over $\s$ we derive the level-matching condition
\bea\la{LMB}
{\cal V} = \int_0^{2\pi} {d\s\ov 2\pi}\left( p_Mx_M' -
\frac{i}{4}P_+\,\str\left(\Sigma_+\chi\chi'\right)
+\frac{1}{2}g_N\bp_M\,\str\left(\left[\S_N,\Sigma_M\right] B_\s\right) \right) =0\,,~~~~~~
\eea
that should be imposed on physical string states.

\subsection{Solving the the Virasoro constraint $C_2$: 
Str$\left(\bp^2 + \l \left(A^{(2)}_1\right)^2\right) = 0$}
Since $x_-'$ does not depend on $\bp_-$, the Virasoro constraint $C_1=0$ can be easily 
solved to find $\bp_-$. The solution has in fact the same form as in bosonic case.
\bea\la{l14}
&&\str\left(\bp^2 + \l\, \left(A^{(2)}_1\right)^2\right) = 
\bp_+\bp_- + \bp_M^2+ \l\, \str\left(\left(A^{(2)}_1\right)^2\right) \\\nonumber
&&={1\ov G_+}\left(P_+ +G_-\bp_-\right)\bp_- 
+ \bp_M^2+ \l\,\str\left( \left(A^{(2)}_1\right)^2\right) = 0\,.
\eea
Let us denote 
\bea
{\cal A}^2\equiv \str\left(\left(A^{(2)}_1\right)^2\right)
\eea
to simplify the notations. Then the solution to this constraint is 
\bea
\label{l15}
\bp_- = - {2 G_+(\bp_M^2 + \l\, {\cal A}^2)\ov P_+ + \sqrt{P_+^2 - 4G_+G_-(\bp_M^2 + \l\, 
{\cal A}^2)}}\,.
\eea

\subsection{Preliminary form of the gauge-fixed Lagrangian}
The gauge-fixed Lagrangian can be now written in the form
\bea
\label{LgfB}
&&\L_{gf} = p_M\dot{x}_M - 
\frac{iP_+}{4}\str\left(\Sigma_+\chi\pa_\tau\chi\right)
+\frac{1}{2}g_N\bp_M\str\left(\left[\S_N,\Sigma_M\right] B_\tau\right)+ \L_{WZ}+ {\bf p_-}
\,.~~~~~~~
\eea
Here we should use the formulas for $\bp_-$, $x'_-$ and ${\bf p_-}$ to express everything in terms 
of physical fields.

\subsection{Simplifying ${\cal A}^2$}
To compute ${\cal A}^2$ we use the following formula
\bea \nonumber
A^{(2)}_1 = -\left({i\ov 4} x_-'\left(\Sigma_-\,g^2+g^2\,\Sigma_-\right) 
+ {1\ov 2}\left(g^{-1}B_1 g -g K_8B_1^tK_8g^{-1}\right)  
+ {1\ov 2}\left(g^{-1}g'+g'g^{-1}\right) \right)\, .\\\la{l17}
\eea
We see that ${\cal A}^2$ is given by the sum of three terms: i) quadratic in $x_-'$, ii) 
linear in $x_-'$, and iii) independent of $x_-'$.

\subsubsection{ Computing the term quadratic in $x_-'$}

This term is equal to 
\bea\la{l18}
&&-{x_-'^2\ov 16}\str\left[\left(\Sigma_-\,g^2+g^2\,\Sigma_-\right) 
\left(\Sigma_-\,g^2+g^2\,\Sigma_-\right)\right] = 
 -4x_-'^2\,G_+G_-\, ,
\eea
where we took into account that $\Sigma_-\,g = g^{-1}\Sigma_-\,, \ \Sigma_-^2 =I_8$, and 
the formula
\bea\la{l21}
g^4 + g^{-4} = 2(G_{tt}+G_{\p\p}-1)I_8 + 8 G_+ G_-\Sigma_8\, .
\eea

\subsubsection{Computing the term linear in $x_-'$}

The linear term is equal to 
\bea\nonumber
&&{i\ov 4}x_-'\str\left[\left(\Sigma_-\,g^2+g^2\,\Sigma_-\right) \left(g^{-1}B_1 g 
-g K_8B_1^tK_8g^{-1}+g^{-1}g'+g'g^{-1}\right)\right]~~~~~~~~~~\\\la{l19}
&&~~~~~~~~~~~~~~~~~~~~~~~~~~~~~~~=2i x_-'G_+ G_-\str\left( \Sigma_+ \chi\chi'\right)\ ,
\eea
where we used that $\str \Sigma_- g^n g' = 0$ for any $n$, and eq.(\ref{l21}).

\subsubsection{Simplifying the term independent of $x_-'$}

The term independent of $x_-'$ is
\bea\nonumber
&&{1\ov 4}\str\left[\left(g^{-1}B_1 g -g K_8B_1^tK_8g^{-1}+g^{-1}g'
+g'g^{-1}\right)^2\right]\\\nonumber
 =&&{1\ov 4}\str\left[\left(g^{-1}B_1 g -g K_8B_1^tK_8g^{-1}\right)^2\right] + 
{1\ov 2}\str\left[\left(g^{-1}B_1 g -g K_8B_1^tK_8g^{-1}\right)
\left(g^{-1}g'+g'g^{-1}\right)\right]
\\\la{l23}
 +&&{1\ov 4}\str\left[\left(g^{-1}g'+g'g^{-1}\right)^2\right]
\eea
The last term independent of $B_1$ in (\ref{l23}) can be easily computed
\bea\la{l24}
{1\ov 4}\str\left[\left(g^{-1}g'+g'g^{-1}\right)^2\right] = 
{z_a'^2\ov (1-{z^2\ov 4})^2} + {y_s'^2\ov (1+{y^2\ov 4})^2}\ .
\eea
The first term in (\ref{l23}) can be cast in the form
\bea\la{l25}
{1\ov 4}\str\left[\left(g^{-1}B_1 g -g K_8B_1^tK_8g^{-1}\right)^2\right] = 
{1\ov 2}\str B_1^2 - {1\ov 2}\str\left(g^{-2}B_1 g^2K_8B_1^tK_8\right)\ .
\eea
Note that this term is of fourth order in fermions.
The second term in (\ref{l25}) can be written in a more explicit form
\bea\nonumber
- {1\ov 2}\str\left(g^{-2}B_1 g^2K_8B_1^tK_8\right)&=& 
-{1\ov 2}(G_+^2+G_-^2)\str\left(B_1 K_8B_1^tK_8\right)-
G_+G_-\str\left(\S_8B_1 K_8B_1^tK_8\right)\\\la{B1}
&&~~~~~~~~~~~~~~~+
{1\ov 2}G_MG_N\str\left(\S_MB_1 \S_NK_8B_1^tK_8\right)\,.
\eea

The second term in (\ref{l23}) can be cast in the form
\bea\la{l26}
&&
{1\ov 2}\str\left[\left(g^{-1}B_1 g -g K_8B_1^tK_8g^{-1}\right)
\left(g^{-1}g'+g'g^{-1}\right)\right]= 
\str\left[B_1\left(g^2\right)'g^{-2}\right]\\\nonumber
&&=(G_+G_-)'\str (\S_8B_1) 
- G_M' G_N \str (\S_M\S_NB_1)= - {1\ov 2}G_M' G_N \str \left(\left[\S_M,\S_N\right]B_1\right)
\, ,
\eea
where we took into account that $2G_+G_-\S_8 - G_MG_N\S_M\S_N+(G_+^2+G_-^2-1)I_8=0$.

\subsubsection{Final form of ${\cal A}^2$}

Collecting the pieces together we get ${\cal A}^2$
\bea \la{l27}
{\cal A}^2&=&\str\left[\left(A^{(2)}_1\right)^2\right] \\\nonumber
&=& -4x_-'^2\,G_+G_- +2i x_-'G_+ G_-\str\left( \Sigma_+ \chi\chi'\right) 
+{z_a'^2\ov (1-{z^2\ov 4})^2} + {y_s'^2\ov (1+{y^2\ov 4})^2}\\\nonumber
&-& {1\ov 2}G_M' G_N \str \left(\left[\S_M,\S_N\right]B_1\right)
+{1\ov 2}\str B_1^2 - {1\ov 2}\str\left(g^{-2}B_1 g^2K_8B_1^tK_8\right)\\\nonumber
&=& -4x_-'^2\,G_+G_- +2i x_-'G_+ G_-\str\left( \Sigma_+ \chi\chi'\right) 
+{z_a'^2\ov (1-{z^2\ov 4})^2} + {y_s'^2\ov (1+{y^2\ov 4})^2}\\\nonumber
&-& {1\ov 2}G_M' G_N \str \left(\left[\S_M,\S_N\right]B_1\right)
+{1\ov 2}\str B_1^2-{1\ov 2}(G_+^2+G_-^2)\str\left(B_1 K_8B_1^tK_8\right)\\\nonumber
&-&G_+G_-\str\left(\S_8B_1 K_8B_1^tK_8\right)+
{1\ov 2}G_MG_N\str\left(\S_MB_1 \S_NK_8B_1^tK_8\right)\,.
\eea
These formulas allow us to compute all necessary terms very efficiently.

\subsection{Simplifying the Wess-Zumino term}
By using the decomposition (\ref{BF}), the odd components of $A_\a$ can be written in the form
\bea\la{Aoddt}
A_\tau^{odd} &=& -ig^{-1}(x)\S_+\chi\sqrt{1+\chi^2} g(x) - g^{-1}(x)F_\tau g(x)\,,\\
\la{Aodds}
A_\s^{odd} &=& - g^{-1}(x)F_\s g(x)\,.
\eea
Then the Wess-Zumino term can be written as a sum of the two terms
\bea\label{WZ2}
&&\L_{WZ} = -\kappa{\sqrt{\lambda}\over 2}\epsilon^{\a\b}\str A_\a^{(1)} A_\b^{(3)} \\\nonumber
&&=i\kappa{\sqrt{\lambda}\over 2}\str\, F_\tau g(x)^2 \widetilde{K}_8 F_\s^t K_8 g(x)^{-2}-
\kappa{\sqrt{\lambda}\over 2}\str\,\S_+\chi\sqrt{1+\chi^2} g(x)^2 \widetilde{K}_8 F_\s^t K_8 g(x)^{-2}
\,.
\eea
It is clear that the second term in (\ref{WZ2}) represents the additional contribution of the Wess-Zumino term to the momentum $p_-$ canonically-conjugate to $x_+$.

Both terms can be written in the following more explicit form by using 
the expression (\ref{gg}) for $g^2$
\bea\la{wzts}
i\kappa{\sqrt{\lambda}\over 2}\str\, F_\tau g(x)^2 \widetilde{K}_8 F_\s^t K_8 g(x)^{-2}
&=& i\kappa{\sqrt{\lambda}\over 2}(G_+^2-G_-^2)\str\left( F_\tau\widetilde{K}_8
F_\s^t K_8\right)
\\
\nonumber
&-&i\kappa{\sqrt{\lambda}\over 2} G_MG_N\str\left(\S_N F_\tau\S_M \widetilde{K}_8
F_\s^tK_8\right)\,,
\eea
\bea\la{wzs}
&&-
\kappa{\sqrt{\lambda}\over 2}\str\,\S_+\chi\sqrt{1+\chi^2} g(x)^2 \widetilde{K}_8 F_\s^t K_8 g(x)^{-2}
\\
\nonumber
&&~~~~~~~~~~~~~~~~~~~~~~~~~~= -
\kappa{\sqrt{\lambda}\over 2}(G_+^2-G_-^2)\str\left(\S_+ \chi\sqrt{1+\chi^2}\widetilde{K}_8F_\s^t
K_8\right) 
\\
\nonumber
&&~~~~~~~~~~~~~~~~~~~~~~~~~~~~~- 
\kappa{\sqrt{\lambda}\over 2}G_MG_N\str\left(\S_+\S_N\chi\sqrt{1+\chi^2} \S_M \widetilde{K}_8
F_\s^t K_8\right)\,.
\eea
The Wess-Zumino term is given by the sum of the terms in (\ref{wzts}) and (\ref{wzs}).

\subsection{Final form of the gauge-fixed Lagrangian}
It is useful to single out a kinetic term and the density of the Hamiltonian from 
the gauge-fixed Lagrangian (\ref{LgfB}), and write it in the form
\bea
\label{LgfB2}
&&\L_{gf} = \L_{kin} - \H
\,,~~~~~~~
\eea
where 
\bea
\label{LkinB}
\L_{kin} &=& p_M\dot{x}_M - 
\frac{iP_+}{4}\str\left(\Sigma_+\chi\pa_\tau\chi\right)
+\frac{1}{2}g_N\bp_M\str\left(\left[\S_N,\Sigma_M\right] B_\tau\right)\\
\nonumber
&+& 
i\kappa{\sqrt{\lambda}\over 2}(G_+^2-G_-^2)\str\left( F_\tau\widetilde{K}_8
F_\s^t K_8\right)
-i\kappa{\sqrt{\lambda}\over 2} G_MG_N\str\left(\S_N F_\tau\S_M \widetilde{K}_8
F_\s^tK_8\right)
\,.~~~~~~~
\eea
\bea
\label{HB}
&&\H = - {\bf p_-} +
\kappa{\sqrt{\lambda}\over 2}(G_+^2-G_-^2)\str\left(\S_+ \chi\sqrt{1+\chi^2}\widetilde{K}_8F_\s^t
K_8\right) 
\\
\nonumber
&&~~~~~~~~~~~~~~~~~~~~~~~~~~~~~+
\kappa{\sqrt{\lambda}\over 2}G_MG_N\str\left(\S_+\S_N\chi\sqrt{1+\chi^2} \S_M \widetilde{K}_8
F_\s^t K_8\right)
\,.~~~~~~~
\eea
Here ${\bf p_-}$ is given by (\ref{pm1B}), and we should use 
the formulas (\ref{l15}), (\ref{l13}) and (\ref{l27}) 
for $\bp_-$, $x'_-$ and ${\cal A}^2$ to express everything in terms 
of physical fields. 

\section{Deriving the quartic Hamiltonian}
In this appendix we derive the various forms of the quartic Hamiltonian 
used in the paper. 
\subsection{Redefining fermions}
The kinetic part (\ref{LkinB}) of the gauge-fixed Lagrangian (\ref{LgfB2}) 
up to the quartic order in fermions can be written  in the form (\ref{Lkin2}) where 
$\Phi(p,x,\chi)$ is given by
\bea
\la{redfermn}
\Phi(p,x,\chi)= &-&{1\ov p_+} {i\ov 2}g_N\bp_M \S_+ \left[\chi,\left[\S_N,\S_M\right]\right]\\\nonumber
&+& {1\ov p_+} \kappa\sqrt{\l}\S_+\left[ 
-{1\ov 2}\widetilde{K}_8(\chi\chi'\chi)^t K_8
-{1\ov 2}\chi\widetilde{K}_8\chi'^t K_8\chi
\right. \\\nonumber
&~&~~~~~~~~+\left(G_+^2-G_-^2-1\right)\widetilde{K}_8
\chi'^t K_8 - \left.G_MG_N\S_M \widetilde{K}_8
 \chi'^tK_8\S_N \right]\,.
\eea
Expanding the bosonic fields, and keeping only terms quartic in fields, we get
\bea
\la{redfer2}
\Phi&=&-{1\ov p_+} {i\ov 4}x_Np_M \S_+ \left[\chi,\left[\S_N,\S_M\right]\right]\\\nonumber
&+& {1\ov p_+} \kappa\sqrt{\l}\S_+\left[ 
-{1\ov 2}\widetilde{K}_8(\chi\chi'\chi)^t K_8
-{1\ov 2}\chi\widetilde{K}_8\chi'^t K_8\chi
\right. \\\nonumber
&~&~~~~~~~~+{1\ov 2}\left(z^2-y^2\right)\widetilde{K}_8
\chi'^t K_8 - \left.x_Mx_N\S_M \widetilde{K}_8
 \chi'^tK_8\S_N \right]\,.
\eea
Then the redefinition of $\chi$ is 
\bea\la{chiredC}
\chi\rightarrow \chi + \Phi(p,x,\chi)\,.
\eea
These shifts of fermions produce additional quartic terms in 
the Hamiltonian but all these terms come only from its fermionic quadratic part
\bea
\H_2^{ferm} =  {\kappa\ov 2}\sqrt{\l}
\str \left(\S_+ \chi\widetilde{K}_8\chi'^t K_8\right) + {p_+\ov 4}\str\,\chi^2\ .\la{Hfermquad}
\eea
The additional quartic terms in the Hamiltonian are equal to
\bea
H_{add}=\kappa\sqrt{\l}
\str \left(\S_+ \Phi\widetilde{K}_8\chi'^t K_8\right) + {p_+\ov 2}\str\,\Phi\chi\ .\la{Hadd}
\eea
The first term in (\ref{Hadd}) gives
\bea
\nonumber
&&\kappa\sqrt{\l}
\str \left(\S_+ \Phi\widetilde{K}_8\chi'^t K_8\right) =
{i\kappa\sqrt{\l}\ov 8p_+}(x_Np_M)'\str\left( 
\left[\S_N,\Sigma_M\right]\left(\widetilde{K}_8\chi^t K_8\chi - \chi\widetilde{K}_8\chi^t K_8\right) 
\right)\\\nonumber
&&~~~~~~~~~~~-{\l\ov 2p_+}\str\left(\chi\chi'\chi\chi'\right) 
-{\l\ov 2p_+}\str\left(\chi\widetilde{K}_8\chi'^t K_8 \chi\widetilde{K}_8\chi'^t K_8 \right) \\
&&~~~~~~~~~~~+{\l\ov 2p_+}(z^2-y^2)\str\left(\chi'\chi'\right) 
-{\l\ov p_+}x_Mx_N\str\left(\S_M\chi'\S_N\chi'\right)\,.\label{Ser1}
\eea
Computing the second term we get
\bea\la{add2}
{p_+\ov 2}\str\,\Phi\chi &=& \frac{i}{4}x_Np_M\str\S_+ \left[\S_N,\Sigma_M\right]\,\chi^2 
-{\kappa\ov 2}\sqrt{\l}\str\Big(
\frac{1}{2}(z^2-y^2)\S_+\chi\widetilde{K}_8\chi'^t K_8 
\\\nonumber
&+&{1\ov 2}\S_+\chi^3\widetilde{K}_8\chi'^t K_8  + {1\ov 2}\S_+\chi\chi'\chi\widetilde{K}_8\chi^t K_8
+ x_Mx_N \S_+\S_N\chi\S_M\widetilde{K}_8
\chi'^t K_8\Big)\,.
\eea
We will see that the first term cancels the same term coming from $p_-$, and
the terms with $\kappa$ just cancel all quartic terms in $\H_{WZ}$ (\ref{Hwz}).

\subsection{${\bf -p_-}$ up to the quartic order}
$-{\bf p_-}$ (\ref{pm1B}) is given by
\bea\nonumber
-{\bf p_- }
&=&{G_-\ov G_+}P_+ - {G_+^2-G_-^2\ov G_+}\bp_-+
\frac{P_+}{4}\str\left( \chi^2\right)\\\la{pm1}
&-&
 \frac{i}{2}g_N\bp_M\str\left(\left[\S_N,\Sigma_M\right]\,\chi^2\left(
\S_+g_+  -\S_-g_- \right) \right) 
\eea

\subsubsection{\ ${G_-\ov G_+}P_+$ up to the sixth order term}

\bea
{G_-\ov G_+}P_+\approx \frac{P_+}{4}  \left(y^2+z^2\right)
-\frac{1}{64} P_+ y^2 z^2 \left(y^2+z^2\right)
\eea
After rescaling (\ref{resc}) it takes the form
\bea
{G_-\ov G_+}P_+\approx \frac{1}{2} \left(y^2+z^2\right)
-\frac{1}{8P_+^2}  y^2 z^2 \left(y^2+z^2\right)
\eea
We see that it does not have a quartic term.

\subsubsection{ ${G_+^2-G_-^2\ov G_+}\bp_-$ up to the quartic order}

Since $G_-^2$ is of the quartic order and $\bp_-$ is of the quadratic order, we get
\bea 
-{G_+^2-G_-^2\ov G_+}\bp_-\approx -G_+\bp_- \approx {1\ov P_+} G_+^2(\bp_M^2 + \l {\cal A}^2)
\eea 
where (see (\ref{l27}))
\bea
{\cal A}^2 \approx&& {z'^2\ov (1-{z^2\ov 4})^2} + {y'^2\ov (1+{y^2\ov 4})^2}\\\nonumber
&+&{1\ov 2}\str B_1^2 - {1\ov 2}\str\left(B_1 K_8B_1^tK_8\right) + 
{1\ov 4}x_M' x_N \str \left(\left[\S_M,\S_N\right](\chi\chi' 
-\chi'\chi)\right)
\eea
We also have
\bea
\str B_1^2 &=& {1\ov 2}\str \left(\chi\chi' \chi\chi'
-\chi^2\chi'^2\right)\\
\str\left(B_1 K_8B_1^tK_8\right)&=&{1\ov 4}\str 
\left((\chi\chi'-\chi' \chi)K_8(\chi\chi'-\chi' \chi)^tK_8\right)
\eea
Then we get 
\bea
{\cal A}^2 \approx&& z'^2 + y'^2 + {1\ov 2} z'^2 z^2 -{1\ov 2} y'^2 y^2 \\\nonumber
&+& {1\ov 4}x_M' x_N \str \left(\left[\S_M,\S_N\right](\chi\chi' -\chi'\chi)\right) \\\nonumber
&+&{1\ov 4}\str\left(\chi\chi'\chi\chi' - \chi^2\chi'^2\right) -
{1\ov 8}\str\left((\chi\chi'-\chi' \chi)K_8(\chi\chi'-\chi' \chi)^tK_8\right)\, .
\eea
The expansion of $G_+^2$ gives
\bea
G_+^2\approx 1+ \frac{1}{2} \left(z^2-y^2\right)\, ,
\eea
so we have
\bea 
-{G_+^2-G_-^2\ov G_+}\bp_-\approx &&{1\ov P_+}\Big(
 p_z^2+p_y^2+ \l(z'^2 + y'^2) \\\nonumber
&+& \frac{1}{2}( p_y^2z^2-p_z^2y^2) + \frac{\l}{2}(y'^2z^2-z'^2y^2)+
\l(z'^2 z^2 - y'^2 y^2)\\\nonumber
&+& \frac{1}{4} \l x_M' x_N \str \left(\left[\S_M,\S_N\right](\chi\chi' -\chi'\chi )\right)
\\\nonumber
 &+&{\l\ov 4}\str\left(\chi\chi'\chi\chi' - \chi^2\chi'^2\right) -
{\l\ov 8}\str\left((\chi\chi'-\chi' \chi)K_8(\chi\chi'-\chi' \chi)^tK_8\right)
 \Big)\, ,
\eea 

\subsubsection{The last term of  $-{\bf p_-}$ up to the quartic order}
For the last term in (\ref{pm1}) we find
\bea
-\frac{i}{2}g_N\bp_M\str\left[\S_N,\Sigma_M\right]\,\chi^2
\left(\S_+g_+-\S_-g_-\right)\approx 
-\frac{i}{4}x_N p_M\str\S_+\left[\S_N,\Sigma_M\right]\,\chi^2\, .
\eea
We see that this term is canceled by the first term in (\ref{add2}). 

\subsubsection{The final result for ${\bf -p_-}$ up to the quartic order}

Summing up the terms we get for $-{\bf p_-}$ 
\bea
\nonumber
-{\bf p_-} \approx &&\frac{1}{4} P_+ \left(y^2+z^2+\str\left( \chi^2\right)\right)+{1\ov P_+}\Big(
 p_z^2+p_y^2+ \l(z'^2 + y'^2) \\\nonumber
&+& \frac{1}{2}( p_y^2z^2-p_z^2y^2) + \frac{\l}{2}(y'^2z^2-z'^2y^2)+
\l(z'^2 z^2 - y'^2 y^2)\\\nonumber
&+& {\l\ov 4}x_M' x_N \str \left(\left[\S_M,\S_N\right](\chi\chi' 
- \chi'\chi)\right)\\\nonumber
 &+&{\l\ov 4}\str\left(\chi\chi'\chi\chi' - \chi^2\chi'^2\right) -
{\l\ov 8}\str\left((\chi\chi'-\chi' \chi)K_8(\chi\chi'-\chi' \chi)^tK_8\right)
 \Big)\\\la{pm3}
&-&\frac{i}{4}x_N p_M\str\S_+\left[\S_N,\Sigma_M\right]\,\chi^2\, ,
\eea 
After rescaling according to (\ref{resc}) this takes the form
\bea
\nonumber
-{\bf p_-} \approx 
&&\frac{1}{2}\left(p_z^2+p_y^2+ z^2+y^2+\tl(z'^2 + y'^2)+\str\left( \chi^2\right)\right)\\\nonumber
&+& {1\ov 2P_+}\Big(p_y^2z^2-p_z^2y^2 + \tl(y'^2z^2-z'^2y^2)+
2\tl(z'^2 z^2 - y'^2 y^2)\\\nonumber
&+& {\tl\ov 2}x_M' x_N \str \left(\left[\S_M,\S_N\right](\chi\chi' - 
\chi'\chi)\right) 
\\\nonumber
&+&{\tl\ov 2}\str\left(\chi\chi'\chi\chi' - \chi^2\chi'^2\right) -
{\tl\ov 4}\str\left((\chi\chi'-\chi' \chi)K_8(\chi\chi'-\chi' \chi)^tK_8\right)\\\nonumber
\la{pm4}
&-&ix_N p_M\str\S_+\left[\S_N,\Sigma_M\right]\,\chi^2 \Big)\, .
\eea

\subsection{Contribution of the WZ term to the Hamiltonian}
Contribution of the Wess-Zumino term to the Hamiltonian (\ref{HB})  is given by
\bea
H_{WZ}
&=&{\kappa\ov 2}\sqrt{\l}\str\Big( (G_+^2-G_-^2)\S_+\chi\sqrt{1+\chi^2}\widetilde{K}_8
F_1^t K_8 \\\nonumber
&~&~~~~~~~~~~~~+ G_MG_N \S_+\S_N\chi\sqrt{1+\chi^2}\S_M\widetilde{K}_8
F_1^t K_8\Big)\,.
\eea
Up to the quartic order we get
\bea\la{Hwz1}
H_{WZ} 
&=&{\kappa\ov 2}\sqrt{\l}\str\Big(\S_+\chi\widetilde{K}_8
\chi'^t K_8 +
\frac{1}{2}(z^2-y^2)\S_+\chi\widetilde{K}_8\chi'^t K_8 
\\\nonumber
&+&{1\ov 2}\S_+\chi^3\widetilde{K}_8\chi'^t K_8  + {1\ov 2}\S_+\chi\chi'\chi\widetilde{K}_8\chi^t K_8
+ x_Mx_N \S_+\S_N\chi\S_M\widetilde{K}_8
\chi'^t K_8\Big)\,.
\eea
We see that it is exactly canceled by the contribution (\ref{add2}) 
coming from the fermion shift (\ref{chiredC}). 

\subsection{Quartic Hamiltonian in terms of $\chi$}
Summing up all the contributions we get the quartic Hamiltonian
\bea\nonumber
\H_4&=&{1\ov 2P_+}\Big[ p_y^2z^2-p_z^2y^2 + \tl (y'^2z^2-z'^2y^2)+
2\tl (z'^2 z^2 - y'^2 y^2)\\\nonumber
 &-&\tl\str\left({1\ov 2}\chi\chi'\chi\chi' + \chi^2\chi'^2+
{1\ov 4}(\chi\chi'-\chi' \chi)K_8(\chi\chi'-\chi' \chi)^tK_8
 +
\chi\widetilde{K}_8\chi'^t K_8 \chi\widetilde{K}_8\chi'^t K_8 \right) \\\nonumber
&+& \tl\str\left( (z^2-y^2)\chi'\chi'
+{1\ov 2} x_M' x_N \left[\S_M,\S_N\right](\chi\chi'- \chi'\chi)
-2 x_Mx_N\S_M\chi'\S_N\chi'\right) 
\\\la{H41C}
&+&{i\kappa\sqrt{\tl}\ov 4}(x_Np_M)'\str\left( 
\left[\S_N,\Sigma_M\right]\left(\widetilde{K}_8\chi^t K_8\chi - \chi\widetilde{K}_8\chi^t K_8\right) 
\right) \Big]  \, .
\eea 
The bosonic part of the Hamiltonian can be further simplified 
if we consider the point-particle reduction of the Hamiltonian, that is  
if we assume the fields to be independent of $\s$, we get
\bea\la{pointp}
\H_{particle}= {p_M^2\ov 2}  + {x_M^2\ov 2} + 
{1\ov 2}\tr\left( \etad\eta +\td\theta\right) + {1\ov 2P_+}(p_y^2z^2-p_z^2y^2)\,.
\eea 
We see first of all that the fermionic part of the Hamiltonian is just given by the quadratic
term. Moreover, the bosonic quartic term can be removed from the Hamiltonian (\ref{pointp}) 
by means of the canonical transformation generated by 
\bea\la{cantran}
V = {1\ov 2P_+}(p_y y\, z^2-p_z z\, y^2)\,.
\eea
Let us recall that given a generating function $V(p,x)$, the canonical transformation of an arbitrary function $f(p,x)$ of the phase space can be written in the form
\bea\la{canV}
f(p,x)\rightarrow \tilde{f}(p,x) &=&  f(p,x) + \{V(p,x),f(p,x)\} + {1\ov 2}\{V,\{V,f\}\} 
+\cdots ~~~~~~
\\\nonumber &=& f+ \sum_{n=1}^\infty {1\ov n!} \underbrace{\{V,\{V,\cdots \{}_n V, f\}\cdots \}\,,
\eea
where the Poisson bracket is defined as
\bea
\{V(p,x),f(p,x)\} = {\pa V(p,x)\ov \pa p_M} {\pa f(p,x)\ov \pa x_M} - 
{\pa V(p,x)\ov \pa x_M} {\pa f(p,x)\ov \pa p_M} \,,
\eea
that is
\bea
\{p_M, x_N\} = \delta_{MN}\,.
\eea
In quantum theory the canonical transformation corresponds to a unitary transformation generated by 
the operator $U$
\bea
U = e^{i V}\,,\quad f(p,x)\rightarrow \tilde{f}(p,x) = U f(p,x)U^\dagger\,,\quad [p,x] = {1\ov i} \,.
\eea
The canonical transformation (\ref{cantran}) can be easily  lifted to $\s$-dependent fields
\bea\la{cantran2}
V = {1\ov 2P_+}\int_0^{2\pi} {d\s\ov 2\pi}(p_y y\, z^2-p_z z\, y^2)\,.
\eea
Then by using (\ref{canV}) one can easily find that the bosonic part of the 
canonically-transformed Hamiltonian takes the form
(\ref{Hbb})
\bea\la{Hbb2}
\H_{bb} 
={\tl\ov p_+}
\left(y'^2z^2 - y^2z'^2 + z^2z'^2 -y^2 y'^2 \right),~~~~
\eea
and we get (\ref{H41}).

The canonical transformation  (\ref{canV}) 
removes {\it all} nonderivative terms from the Hamiltonian. One can show that the sixth order 
nonderivative terms can be also removed by a canonical transformation. This is in accord with 
the observation that already the quadratic particle Hamiltonian reproduces the spectrum of type IIB
supergravity on $\ads$.

\section{Hamiltonian in terms of $\eta$ and $\theta$}
In this appendix we use the decomposition (\ref{chiT}) for $\chi$ to 
express (\ref{H41}) in terms of $\eta$ and $\theta$. 

\medskip

{\bf Quadratic Hamiltonian}

\medskip

We first use
\bea\nonumber
\str(\chi^2) &=& 2\tr\left(\T\T_*\right) = - 2\tr\left(\S\Td\T\right)\,,\\\nonumber
\str\left(\S_+ \chi\widetilde{K}_8\chi'^t K_8 \right) &=&
-\tr\left(\S \T K \T'^t K \right) -\tr\left(\S \Td K \T'^{\dagger,t} K \right)\,,
\eea
to write the quadratic Hamiltonian in the form
\bea\nonumber
\H_2 = {1\ov 2} p_M^2 + {1\ov 2}x_M^2 +{\tl\ov 2}x_M'^2  - {\kappa\ov 2}\sqrt{\tl}\left(
\tr\left(\S \T K \T'^t K \right) 
+\tr\left(\S \T^\dagger K \T'^{\dagger,t}K \right)\right) - \tr\left(\S\T^\dagger\T\right)\, .
\eea
Then by using 
\bea\nonumber
\tr\S\Td\T &=& -{1\ov 2}\tr\left(  \etad\eta +\td\theta\right)\\\nonumber
\tr\S\T K\T'^t K&=& {1\ov 2}\tr\left( - \eta\eta' +\td\tpd\right)\,,\quad
\tr\S\Td K\Tpdt K= {1\ov 2}\tr\left( \etad\etapd -\theta\theta'\right)\,,
\eea
we obtain (\ref{Hquadr4}).

\medskip

{\bf Quartic Hamiltonian}

\medskip

We first use the following relations
\bea\nonumber
&&\str\left(\chi'\chi'\right) = -2\tr\left(\S\T'  \T'^\dagger \right)\,,\quad
\str\left(\chi\chi'\chi\chi'\right)
=\tr\left(\T\Tpd\T\Tpd-\Td\T' \Td\T'\right)\,,~~~~~\\\nonumber
&&\str\left(\chi^2\chi'^2\right)
=\tr\left(\T\Td\T'\Tpd-\Td\T \Tpd\T'\right)\,,~~~~~\\\nonumber
&&\str\left((\chi\chi' - \chi'\chi)K_8(\chi\chi' -\chi'\chi)^t K_8\right)\\\nonumber
&&~~~=\tr\left((\T\Tpd - \T'\Td) K(\T\Tpd -\T'\Td)^t K-(\Td\T'- \Tpd\T) K(\Td\T' - \Tpd\T)^t K\right)
\,,~~~~~\\\nonumber
&&\str\left(\chi\widetilde{K}_8\chi'^t K_8 \chi\widetilde{K}_8\chi'^t K_8 \right)=
\tr\left( K\Tpdt K\Td K\Tpdt K\Td\right) -\tr\left( K\T'^tK\T K\T'^tK\T\right)\,,\\\nonumber
&&x_Mx_N\str\left(\S_M\chi'\S_N\chi'\right) = -
2iZ_m Y_n\tr\left( \S\G_m\T' \G_n\Tpd\right)\,,\\\nonumber
&&x_M' x_N \str \left(\left[\S_M,\S_N\right](\chi\chi'- \chi'\chi)\right)\\\nonumber  
&&~~~=-Z_m'Z_n\tr\left(\S\left[\G_m,\G_n\right](\T\Tpd - \T'\Td)\right)+
Y_m'Y_n\tr\left(\S\left[\G_m,\G_n\right](\Td\T' -\Tpd\T )\right) \,,\\\nonumber
&&(x_Np_M)'\str\left( 
\left[\S_N,\Sigma_M\right]\left(\widetilde{K}_8\chi^t K_8\chi - \chi\widetilde{K}_8\chi^t K_8\right) 
\right)\\\nonumber 
&&~~~=2(Z_n P^z_m)'\tr\left( \left[\G_n,\G_m\right](K\Tdt K\Td+ \T K\T^tK)\right)  \\\nonumber 
&&~~~~~~~-2(Y_n P^y_m)'\tr\left( \left[\G_n,\G_m\right](K\T^tK\T +\Td K\Tdt K) \right)\,,
\eea
to write the quartic Hamiltonians $\H_{bf}$ and $\H_{ff}$ in the form
\bea\la{HbfD}
\H_{bf} &=& {1\ov 2p_+}\Big[
-2\tl (z^2-y^2)\tr\left(\S\T'  \T'^\dagger \right) +4i\tl Z_m Y_n\tr\left( \S\G_m\T' \G_n\Tpd\right)~~~~~~~\\\nonumber
&~&~~~~~ -{\tl\ov 2}\, \tr\left(
Z_m'Z_n\S\left[\G_m,\G_n\right](\T\Tpd - \T'\Td) - 
Y_m'Y_n\S\left[\G_m,\G_n\right](\Td\T'- \Tpd\T)\right) \\\nonumber
&~&~~~~~ + {i\kappa\ov 2}\sqrt{\tl}\,\tr\left((Z_n P^z_m)'
 \left[\G_n,\G_m\right]( K\Tdt K\Td + \T K\T^tK)\right.\\\nonumber
&~&~~~~~~~~~~~~~~~~~~~~~~\left. -(Y_n P^y_m)'\left[\G_n,\G_m\right]( K\T^tK\T +\Td K\Tdt K)\right)\Big]\, ,
\eea
\bea\la{HffD}
\H_{ff} &=& {1\ov 2p_+}\Big[
-{\tl\ov 2}\tr\left(\T\Tpd\T\Tpd-\Td\T' \Td\T' +\T\Td\T'\Tpd-\Td\T \Tpd\T'\right)\\\nonumber
&-&{\tl\ov 4}
\tr\left((\T\Tpd - \T'\Td) K(\T\Tpd -\T'\Td)^t K-(\Td\T'- \Tpd\T) K(\Td\T' - \Tpd\T)^t K\right)
\\\nonumber
&~&~~~~~ -\tl\,\tr\left( K\Tpdt K\Td K\Tpdt K\Td - K\T'^tK\T K\T'^tK\T\right)
\Big]\, .
\eea
Note that we used 
\bea
z_m\g_m= Z_m\G_m\, ,\quad y_m\g_m= Y_m\G_m  
\, ,\quad p_m^y\g_m= 2P_m^y\G_m \,,\quad p_m^y\g_m= 2P_m^y\G_m\,.
\eea
Then by using 
\bea\nonumber
\tr\S\left(\Tpd\T'\right)&=& -{1\ov 2}\tr\left(\etapd\eta' +\tpd\theta'\right)\,,\\\nonumber
\tr\S\G_m\G_n\left(\T\Tpd -\T'\Td\right)&=& 
\tr\left( \PP_+ \G_m\G_n(\eta\etapd - \eta'\etad)- 
\PP_- \G_m\G_n(\td\theta' -\tpd\theta)\right)\,,~~~~~~~~~~\\\nonumber
\tr\S\G_m\G_n\left(\Td\T' - \Tpd\T\right)&=& \tr\left(- \PP_- \G_m\G_n(\etad\eta' -\etapd\eta) 
+ \PP_+ \G_m\G_n(\theta\tpd -\theta'\td)\right)\,,~~~~~~~~~~\\\nonumber
\tr\S\G_m\T'\G_n\Tpd&=& \tr\left(- \PP_- \G_m\eta'\G_n\theta'+ \PP_+ \G_m\tpd\G_n\etapd \right)\,,
\eea
\bea\nonumber
\tr\left(\left[\G_n,\G_m\right]( K\Tdt K\Td + \T K\T^tK)\right) &=& \\\nonumber
-\tr \PP_+\left[\G_n,\G_m\right]( \etad\etad +\eta\eta) &-&
\tr \PP_-\left[\G_n,\G_m\right]( \td\td +\theta\theta)\,,
\\\nonumber
\tr\left(\left[\G_n,\G_m\right]( K\T^tK\T +\Td K\Tdt K)\right)&=& \\\nonumber
-\tr \PP_-\left[\G_n,\G_m\right]( \etad\etad +\eta\eta) 
&-&\tr \PP_+\left[\G_n,\G_m\right]( \td\td +\theta\theta)\,,
\eea
we obtain (\ref{Hbf2}).

To find $H_{ff}$ we need
\bea\nonumber
&&\tr\left(\T\Tpd\T\Tpd-\Td\T' \Td\T' +\T\Td\T'\Tpd-\Td\T \Tpd\T'\right)\\\nonumber
&&~~~~~= 
\tr\left(\PP_+\eta\etapd\eta\etapd - \PP_- \etad\eta'\etad\eta' + 
\PP_+\eta\etad\eta'\etapd - \PP_- \etad\eta\etapd\eta'\right. \\\nonumber
&&~~~~~- \left.
\PP_+\theta\tpd\theta\tpd + \PP_- \td\theta'\td\theta' -
\PP_+\theta\td\theta'\tpd + \PP_- \td\theta\tpd\theta'\right)\,,\\\nonumber
&&
\tr\left((\T\Tpd - \T'\Td) K(\T\Tpd -\T'\Td)^t K
- (\Td\T'- \Tpd\T) K(\Td\T' - \Tpd\T)^t K  \right)\\\nonumber
&&~~~~~=
\tr\S\left((\eta\etapd - \eta'\etad)(\etapd\eta - \etad\eta')
-(\tpd\theta- \td\theta')(\theta\tpd-\theta'\td )\right)\,,\\\nonumber
&&\tr\left( K\Tpdt K\Td K\Tpdt K\Td - K\T'^tK\T K\T'^tK\T\right)\\
&&~~~~~=\nonumber
\tr\left( \PP_+ \etapd\etad\etapd\etad - \PP_- \eta'\eta\eta'\eta - 
\PP_+ \tpd\td\tpd\td +\PP_-\theta'\theta\theta'\theta\right)\,.
\eea
By using these formulas we find the following expression for $H_{ff}$
\bea
H_{ff} = H_{ff}(\eta) -  H_{ff}(\theta)\, ,
\eea
where 
\bea 
H_{ff}(\eta)&=& {\tl\ov 2p_+}\tr\Big[
-{1\ov 2}\left(\PP_+\eta\etapd\eta\etapd - \PP_- \etad\eta'\etad\eta' + 
\PP_+\eta\etad\eta'\etapd - \PP_- \etad\eta\etapd\eta'\right)~~~~~~~~~~ \\\nonumber
&~&~~~~~~~~~~~-{1\ov 4}\S(\eta\etapd - \eta'\etad)(\etapd\eta - \etad\eta')
-\left(\PP_+ \etapd\etad\etapd\etad - \PP_- \eta'\eta\eta'\eta \right)
\Big]\, .
\eea
It can be cast in the form
\bea \la{hffeta}
H_{ff}(\eta)&=& {\tl\ov 2p_+}\tr\Big[
-{1\ov 2}\S\left(\etapd\eta\etapd\eta + \etad\eta'\etad\eta' + 
\etapd\etad\etapd\etad + \eta'\eta\eta'\eta
\right)~~~~~~~~~~ \\\nonumber
&&~~-{1\ov 4}\left(\eta\etapd\eta\etapd -  \etad\eta'\etad\eta' + 
\eta\etad\eta'\etapd - \etad\eta\etapd\eta' +2\etapd\etad\etapd\etad - 2 \eta'\eta\eta'\eta
\right)
\Big]\, .
\eea
By using the formula 
\bea
\tr \G_a\G_{5-b}\G_c\G_{5-d} = \delta_{ab} \delta_{cd}+ \delta_{ad} \delta_{bc} -
 \delta_{ac} \delta_{bd}\,,     
\eea
one can check that the second line in (\ref{hffeta}) vanishes, and, therefore, $H_{ff}(\eta)$ 
takes the form (\ref{hffeta2}).




\end{document}